\documentclass[useAMS,usenatbib]{mn2e}

\usepackage{graphicx}
\usepackage{amstext}
\usepackage{amsmath}
\usepackage{natbib}
\usepackage{url}
\usepackage{setspace}
\usepackage{tabularx}
\usepackage{mathtools}
\usepackage{multicol}
\usepackage{blindtext}
\usepackage{lscape}
\usepackage{rotating}
\usepackage{caption}
\usepackage{verbatim}
\usepackage{multirow}
\usepackage{color}

\bibpunct[ ]{(}{)}{,}{a}{}{,}     
  
\def	 \meth {{\rm CH_{3}OH}}
\def	 \hmol	{{\rm HN^{13}C}}
\def 	 \hion {{\rm H^{13}CO^{+}}}
\def	 \onetzero {{\rm $J=1\rightarrow0$}}
\def	 \twotone {{\rm $J=2\rightarrow1$}}
\def	 \threettwo {{\rm $J=3\rightarrow2$}}
\def	 \onetzeroj {{\rm $1-0$}}
\def	 \twotonej {{\rm $2-1$ }}

\def     \sol {{\rm M$_\odot$}}
\def     \arcsec {{\rm $^{\prime\prime}$}}
\def     \irdcC {{\rm G028.37+00.07}}
\def     \irdcF {{\rm G034.43+00.24}}
\def     \irdcG {{\rm G034.77-00.55}}
\def     \irdcH {{\rm G035.39-00.33}}

\def     \ltsimm{\mathrel{\spose{\lower 3pt\hbox{$\sim$}}\raise 2.0pt\hbox{$<$}}}
\def     \gtsimm{\mathrel{\spose{\lower 3pt\hbox{$\sim$}}\raise 2.0pt\hbox{$>$}}}

\title[SiO and CH$_3$OH Emission in IRDCs]{Widespread SiO and CH$_3$OH Emission 
    in Filamentary Infrared-Dark Clouds\thanks{Based on observations carried out with the IRAM 30m Telescope under projects 134-12 and 027-13. IRAM is supported by INSU/CNRS (France),
    MPG (Germany) and IGN (Spain).}}  \author[G. Cosentino et al.]
      {G. Cosentino$^{1}$\thanks{E-mail:giuliana.cosentino.15@ucl.ac.uk},
        I. Jim\'{e}nez-Serra$^{2}$, J. D. Henshaw$^{3}$,  P. Caselli$^{4}$, S. Viti$^{1}$, \newauthor A. T. Barnes$^{3,4}$, F. Fontani$^{6}$, J. C. Tan$^{5}$, A. Pon$^{7}$\\
$^{1}$Department of Physics and Astronomy, University College London, Gower Street, London WC1E6BT, UK \\
$^{2}$Astronomy Unit, Queen Mary University of London, Mile End Road, London E1 4NS, UK\\
$^{3}$Max Planck Institute for Astronomy, K\"onigstuhl 17, D-69117 Heidelberg, Germany\\
$^{4}$Max Planck Institute for Extraterrestrial Physics, Giessenbachstrasse 1, 85748 Garching bei M\"unchen, Germany\\
$^{5}$Departments of Astronomy and Physics, University of Florida,Gainesville, FL 32611, USA\\
$^{6}$INAF  Osservatorio Astronomico di Arcetri, Largo E. Fermi 5, 50125 Florence, Italy\\
$^{7}$Department of Physics and Astronomy, The University of Western Ontario, 1151 Richmond Street, London, N6A 3K7, Canada}

\begin{document}

\date{Accepted 2017 November 24. Received 2017 November 24; in original form 2017 June 07}

\pagerange{\pageref{firstpage}--\pageref{lastpage}} \pubyear{2017}

\maketitle
\nocite{*}
\begin{abstract}\label{abstract}\label{firstpage}
Infrared-Dark Clouds (IRDCs) are cold, dense regions of high (optical and infrared) extinction, believed to be the birthplace of high-mass stars and stellar clusters. The physical mechanisms leading to the formation of these IRDCs are not completely understood and it is thus important to study their molecular gas kinematics and chemical content to search for any signature of the IRDCs formation process. Using the 30m-diameter antenna at the Instituto de Radioastronom\'ia Milim\'etrica, we have obtained emission maps of dense gas tracers ($\hion$ and $\hmol$) and typical shock tracers (SiO and $\meth$) toward three IRDCs, \irdcC, \irdcF \space and \irdcG \space (clouds C, F and G, respectively). We have studied the molecular gas kinematics in these clouds and, consistent with previous works toward other IRDCs, the clouds show complex gas kinematics with several velocity-coherent sub-structures separated in velocity space by a few km s$^{-1}$. Correlated with these complex kinematic structures, widespread (parsec-scale) emission of SiO and $\meth$ is present in all the three clouds. For clouds C and F, known to be actively forming stars, widespread SiO and $\meth$ is likely associated with on-going star formation activity. However, for cloud G, which lacks either 8 $\mu$m or 24 $\mu$m sources and 4.5 $\mu$m H$_2$ shock-excited emission, the detected widespread SiO and $\meth$ emission may have originated in a large-scale shock interaction, although a scenario involving a population of low-mass stars driving molecular outflows cannot be fully ruled out.

\end{abstract}


\section{Introduction}\label{introduction}
The process that leads to the formation of high-mass stars (stars with masses $>$ 8 \sol) is still under debate. The difficulty in tracing their evolutionary path lies in their short evolutionary time-scales, their crowded cluster environments, and the large distances at which they are typically found. In addition, their strong feedback (stellar winds, supernovae explosions and UV radiation) largely affects their birth place, making the study of their formation processes even more challenging.   

One way to circumvent the problem of feedback is to study the earliest stages of high-mass star formation represented by infrared-dark clouds  \citep[or IRDCs;][]{simon2006,rathborne2006,peretto2010,butler2012}. IRDCs are cold \citep[T $\leq$ 25 K;][]{pillai2006}, dense (n(H$_2$) $\geq$ 10$^5$ cm$^{-3}$) and highly-extinguished \citep[A$_v$ $\geq$100 mag and N(H$_2$) $\geq$ 10$^{22}$ cm$^{-2}$;][]{butler2009,butler2012} molecular clouds, first observed in extinction against the bright mid-IR Galactic background \citep{perault1996,egan1998}. The morphology ranges globular to very filamentary, with the latter morphology likely denoting the earliest stages in their evolution as predicted by simulations of molecular cloud formation \citep{vanloo2007,hennebelle2008,heitsch2009}. Several scenarios have been proposed to explain the formation of IRDCs, including flow-driven formation, gravitational collapse and cloud-cloud collisions \citep[e.g.][]{hennebelle2008, heitsch2009,tasker2009,vanloo2014,wu2015}. Among them, the cloud-cloud collision scenario proposes that IRDCs can form after the collision of already-molecular clouds, and since the collision is expected to be relatively gentle \citep[with velocities from a few to 10 km s$^{-1}$ as opposed for instance to the flow-driven scenario, e.g.][]{wu2016,wu2017a,wu2017b} signatures of the cloud-cloud collision (which involves a large-scale shock interaction) should be observable in the kinematics and excitation of the molecular line spectra observed toward young IRDCs \citep[e.g][]{wu2015}.

One of the best tracers of shock interactions in star-forming regions is silicon monoxide (or SiO). This molecule is known to be heavily depleted in the quiescent gas of molecular dark clouds \citep[with measured upper limits to the SiO abundance of $\leq$10$^{-12}$;][]{martinpintado1992,jimenezserra2005}, but its abundance is enhanced by up to six orders of magnitude in shocked regions associated with molecular outflows \citep[measured SiO abundances as high as $\sim$10$^{-6}$;][]{martinpintado1992,jimenezserra2005}. SiO is thought to be produced in shocks after the release of Si into the gas phase mainly by the sputtering of dust grains \citep[either from the icy mantles or from the grain cores; see][]{schilke1997,jimenezserra2008}.  
Depending on the velocity of the shock, the SiO emission will show different line profiles: while the SiO gas in high-velocity shocks presents broad line emission red- and blue-shifted by tens of km s$^{-1}$, the SiO emission in low-velocity shocks should be significantly narrower \citep[with linewidths $\leq$1-2 km s$^{-1}$;][]{jimenezserra2009}.

One of the first attempts to detect the signature of a cloud-cloud collision in an IRDC was reported by \cite{jimenezserra2010}. In this work, These authors presented the detection of widespread emission of SiO toward the filamentary IRDC G035.39-00.33 \citep[or cloud H in][]{butler2009}. The detailed analysis of the SiO line profiles across this cloud revealed two different contributions: i) a bright and compact component with broad line emission clearly associated with on-going star formation activity; and ii) a weak, narrow (linewidths $\leq$2 km s$^{-1}$) and extended component detected even toward the most quiescent regions in the IRDC. Among other possibilities, \cite{jimenezserra2010} proposed that the widespread and narrow component of SiO could have been generated by a large-scale cloud-cloud collision related to the origin of the IRDC itself. This cloud indeed shows a complex kinematic structure with several velocity-coherent filaments separated in velocity space by $\sim$3 km s$^{-1}$, which seem to be interacting \citep[see][]{henshaw2013, henshaw2014, jimenezserra2014}. Therefore, the proposed {\it gentle} shock interaction between the velocity-coherent filaments in \irdcH \space may have injected enough Si into the gas phase, yielding narrow SiO line profiles as a distinct chemical signature. 

Since this first study, other works have analyzed the morphology and kinematics of the line emission of SiO in massive molecular clouds, but mainly targeting regions with on-going star formation \citep[see e.g. the W43 and Cygnus X star-forming regions;][]{nguyen2013, duartecabral2014}. More recently, \cite{csengeri2016} performed a spectral line survey of massive clumps located in the Galactic plane, covering different  evolutionary stages from IRDCs to Ultra Compact (UC) HII regions. These authors found that the SiO emission is detected in almost all sources, although the shape of the SiO line profiles clearly varies as a function of evolutionary stage. Indeed, while the most active clumps show line profiles with two clear components (a 20 km s$^{-1}$ wide broad component and a 5-6 km s$^{-1}$ wide narrow component), in the most quiescent sources (mostly associated with IRDCs) only the narrow component is detected. These findings indicate that narrow SiO emission may be a common feature in IRDCs and that it may arise from low-velocity shocks. 
Unfortunately, \cite{csengeri2016} carried out single-pointing observations and therefore, the overall morphology of this narrow SiO emission in other IRDCs remains unknown.

The aim of this work is to extend the study carried out by \cite{jimenezserra2010} in \irdcH \space to other IRDCs known to be at a relatively early stage in their evolution (as measured by their low levels of star formation activity). To this purpose, we have carried out large-scale maps of the emission of SiO and methanol (CH$_3$OH), and of dense gas tracers ($\hion$ and $\hmol$), toward three IRDCs, $\irdcC$, $\irdcF$ and $\irdcG$ \citep[clouds C, F and G respectively, following][]{butler2012}. Like SiO, CH$_3$OH is also known to be enhanced in outflows by several orders of magnitude \citep[e.g. L1157-mm and L1448-mm;][]{bachiller2001,jimenezserra2005}, and therefore it is a good probe of the interaction of low-velocity shocks. The three clouds of our study belong to the sample of ten clouds selected by \cite{butler2009} from the \cite{rathborne2006} catalog. They were selected because they are nearby, massive and show high contrast compared to the surrounding diffuse emission.

The paper is organized as follows. In Sec.~\ref{observations} we provide details about the observations while in Sec.~\ref{method} we describe the tools and procedures adopted for our data analysis. In Sec.~\ref{results}, we present our results on the detection of widespread SiO and CH$_3$OH emission toward $\irdcC$, $\irdcF$ and $\irdcG$ and analyze their kinematics in detail. In Sec.~\ref{abundances}, we calculate the abundances of SiO and CH$_3$OH across the clouds, and in Sec.~\ref{discussion} and Sec.~\ref{conclusion} we discuss our results and present our conclusions. 

\section{Observations}\label{observations}

The \onetzero\, rotational transitions of $\hion$ and $\hmol$, the \twotone\, line of SiO and the \threettwo\, transitions of $\meth$, 
were mapped toward the IRDCs \irdcC, \irdcF \space and \irdcG \space in May and September 2013 using the IRAM (Instituto de Radioastronom\'{\i}a Milim\'etrica) 30m telescope in Pico Velata, Spain. For simplicity, hereafter we will refer to \irdcC\,  as cloud C, \irdcF\space as cloud F and \irdcG\, as cloud G. The map sizes were 264\arcsec $\times$252\arcsec \space for cloud C, 144\arcsec$\times$264\arcsec \space for cloud F and 204\arcsec$\times$240\arcsec \space for cloud G. 
Maps were obtained using the On-The-Fly (OTF) observing mode with central coordinates at
$\alpha(J2000) = 18^h42^m52.3^s$, $\delta(J2000) = -4^{\circ}02^{\prime}26.2$\arcsec \space for cloud C, 
\hbox{$\alpha(J2000) = 18^h53^m18^s$}, $\delta(J2000) = 1^{\circ}27^{\prime}22.1$\arcsec \space for cloud F and $\alpha(J2000) = 18^h56^m45^s$, $\delta(J2000) = 1^{\circ}21^{\prime}45$\arcsec for cloud G. The off-positions were, respectively, \hbox{(-370\arcsec, 30\arcsec)}, \hbox{(-200\arcsec, 0\arcsec)}, and (-240\arcsec, -40\arcsec) for clouds C, F and G and the angular separation in the direction perpendicular to the scanning direction was 6\arcsec. In Table~\ref{tab1}, we report the spectroscopic information of the observed transitions from the CDMS catalog\footnote{See https://www.astro.uni-koeln.de/cdms/catalog}, as well as the beam size and beam efficiency of the IRAM 30m telescope at these frequencies. 

For the observations both the VESPA and FTS spectrometers were used. VESPA provided a spectral resolution of 40 kHz,
corresponding to a velocity resolution of 0.14 km s$^{-1}$ for $\hion$, $\hmol$ and SiO, and of 0.08 km s$^{-1}$ for $\meth$. The FTS spectrometer provided a spectral resolution of 190 kHz, corresponding to velocity resolutions of 0.40 km s$^{-1}$ for $\meth$ and 0.70 km s$^{-1}$ for the other molecules. For our analysis, we used the data collected with the VESPA spectrometer for all the maps except for $\meth$ in clouds F and G, for which the FTS data have been used. The peak intensities were measured in units of antenna temperature, T$^{*}_{A}$, and converted into
main-beam temperature, T$_{mb}$, using a beam and forward efficiencies of of 0.81 and 0.95 for the 3$\,$mm data and 0.73 and 0.93 for the 2$\,$mm spectra, respectively. 

The final data cubes were created using the {\sc CLASS} software within the {\sc GILDAS} package\footnote{See http://www.iram.fr/IRAMFR/GILDAS.}. The maps have a spatial resolution of 30\arcsec \space and a pixel size of 15\arcsec$\times$15\arcsec \space for SiO, $\hion$ and $\hmol$, and
a spatial resolution of 18\arcsec \space and a pixel size of 9\arcsec$\times$9\arcsec \space for $\meth$. For all the datacubes, we used a Gaussian convolving kernel with $\sigma$ = 6\arcsec \space for $\meth$ and $\sigma$ = 10\arcsec \space for all the other tracers. In most cases, the spectra were smoothed in velocity to improve the signal-to-noise ratio of the measured line emission. The final properties of every data cube are listed in Table~\ref{tab2}.

\begin{table*}
\centering 
\caption{Observed molecular transitions, frequencies and beam-sizes of the raw data.}
\begin{tabular}[b]{llcccccc}
\hline
\hline
Molecule & Transition & Frequency& E$_{up}$ & g$_{up}$ & A$_{ul}$ ($\times$10$^{-5}$) & $\theta$& Beam Eff.\\
 &  &  (GHz) & (K) & & s$^{-1}$  & (\arcsec) & \\
\hline
$\hion$ & \onetzeroj            & 86.75429 &4.2 &3 &3.85 &28 &0.81\\
$\hmol$ & \onetzeroj            & 87.09083 &4.2 &3 &2.38 &28 &0.81\\
SiO     & \twotonej             & 86.84699 &6.3 &5 &2.93 &28 &0.81\\
$\meth$ & 3$_{1,2}$-2$_{1,1}$   &145.13186 &35.0 &7 &1.12 &17 &0.73\\
$\meth$ & 3$_{-2,2}$-2$_{-2,1}$ &145.12639 &39.8 &7 &0.69 &17 &0.73\\
$\meth$ & 3$_{0,3}$-2$_{0,2}^{++}$   &145.10315 &13.9 &7 &1.23 &17 &0.73\\
$\meth$ & 3$_{-1,3}$-2$_{-1,2}$ &145.09737 &19.5 &7 &1.10 &17 &0.73\\
$\meth$ & 3$_{0,3}$-2$_{0,2}$   &145.09371 &27.1 &7 &1.23 &17 &0.73\\
\hline
\end{tabular}
\label{tab1}
\end{table*}

\begin{table*}
\centering
\caption{Observed molecular transitions, final beam-sizes and velocity resolution of the data cubes used in our analysis.}
\begin{tabular}[t]{llcccc}
\hline
\hline
        & Molecule  & Beam size  & $\delta$v & $r.m.s.$  & $\delta \nu$ \\
   &  & (\arcsec) & (km s$^{-1}$) & (K) & (kHz) \\ 
\hline
Cloud C & \multicolumn{4}{l}{$\alpha$(J2000)=18:42:52.3, $\delta$(J2000)=-4:02:26.2, map size=264\arcsec$\times$252\arcsec} \\
        &$\hion$ (\onetzeroj) &30 &0.28 &0.010 &80 \\
       	&$\hmol$ (\onetzeroj) &30 &0.28 &0.007 &80 \\
        &SiO (\twotonej)                 &30 &0.56 &0.010 &160 \\
        &$\meth$ (3$_{0,3}$-2$_{0,2}^{++}$ )  &18 &0.32 &0.014 &160 \\
	    &$\meth$ (3$_{-1,3}$-2$_{-1,2}$ )&18 &0.32 &0.014 &160 \\
        &$\meth$ (3$_{0,3}$-2$_{0,2}$ )  &18 &0.32 &0.014 &160 \\ \hline
Cloud F & \multicolumn{4}{l}{$\alpha$(J2000)=18:53:18, $\delta$(J2000)=1:27:22.1, map size=144\arcsec$\times$264\arcsec} \\
        &$\hion$  (\onetzeroj)           &30 &0.14 &0.012 &40  \\
        &$\hmol$  (\onetzeroj)           &30 &0.14 &0.011 &40  \\
        &SiO  (\twotonej)                &30 &0.56 &0.006 &160  \\
        &$\meth$ (3$_{1,2}$-2$_{1,1}$ )  &18 &0.80 &0.005 &380 \\
        &$\meth$ (3$_{-2,2}$-2$_{-2,1}$) &18 &0.80 &0.005 &380 \\
        &$\meth$ (3$_{0,3}$-2$_{0,2}^{++}$ )  &18 &0.80 &0.005 &380 \\
	    &$\meth$ (3$_{-1,3}$-2$_{-1,2}$ )&18 &0.80 &0.005 &380 \\
        &$\meth$ (3$_{0,3}$-2$_{0,2}$ )  &18 &0.80 &0.006 &380 \\ \hline
Cloud G & \multicolumn{4}{l}{$\alpha$(J2000)=18:56:45, $\delta$(J2000)=1:21:45.0, map size=204\arcsec$\times$240\arcsec} \\   
        &$\hion$ (\onetzeroj)             &30 &0.28 &0.011 &80  \\
        &$\hmol$ (\onetzeroj)             &30 &0.28 &0.010 &80  \\
        &SiO (\twotonej)                  &30 &0.28 &0.011 &80  \\
        &$\meth$ (3$_{0,3}$-2$_{0,2}^{++}$ )   &18 &0.40 &0.008 &190  \\
	    &$\meth$ (3$_{-1,3}$-2$_{-1,2}$ ) &18 &0.40 &0.008 &190  \\
        &$\meth$ (3$_{0,3}$-2$_{0,2}$ )   &18 &0.40 &0.008 &190  \\
\hline
\end{tabular}
\label{tab2}
\end{table*}

\section{Analysis Method: SCOUSE}\label{method}
As shown by \cite{henshaw2014, henshaw2016}, if the molecular gas in a cloud has multi-component line profiles, a first moment analysis of the molecular emission only provides the average radial velocity between the different components along the line of sight. Therefore, to study in detail the kinematics of the molecular gas in IRDCs, a multi-Gaussian fitting approach needs to be used in order to isolate the gas motions of the individual velocity components. In our case, we performed multi-gaussian fitting of all the spectra measured in every data cube using the IDL {\sc SCOUSE}\footnote{See https://github.com/jdhenshaw/SCOUSE} code (\textit{Semi-automated multi-COmponent Universal Spectral-line fitting Engine}) recently developed by \cite{henshaw2016}.

{\sc SCOUSE} is a semi-automated procedure developed to fit large amounts of spectra from single-dish and interferometric data cubes in a systematic and efficient way. As a first step, the tool excludes from the analysis the regions where the line peak intensity is lower than a user-provided threshold. Then, it divides the significant remaining region into several smaller areas called \textit{Spatial Averaging Areas} (SAAs), whose sizes are also defined by the user. All the spectra contained in a SAA are spatially averaged and the user is required to manually fit the average spectrum. Finally, SCOUSE uses the best fitting values of the average spectrum as input parameters to perform multi-gaussian fitting for every single spectrum in a SAA. This procedure is repeated for all the identified SAA. In order to make the final fittings physically consistent, the user is also required to provide tolerance levels for the derived peak line intensity, centroid radial velocity and linewidth (FWHM). The {\sc SCOUSE} analysis thus provides information on the peak intensity, centroid radial velocity and FWHM for all velocity components in all emission lines, and on the measured RMS noise level in every spectrum. Other statistical parameters, such as the residual value, $\chi^{2}$, $\tilde{\chi}^{2}$ (the $\chi^{2}$ divided by the number of free parameters) and the Akaike Information Criterion \citep[AIC;][]{akaike1974} are also calculated. All this information has been used to evaluate the distribution in radial velocity and line width of the mapped molecular emission in all the clouds of our sample. For a more detailed discussion on SCOUSE see \cite{henshaw2016}. 

\section{Results}\label{results}
\subsection{The Dense Gas Tracers}\label{dense_gas_emission}
\begin{figure*}
\includegraphics[trim=1cm 2.1cm 2cm 1.7cm,clip=true,scale=0.37]{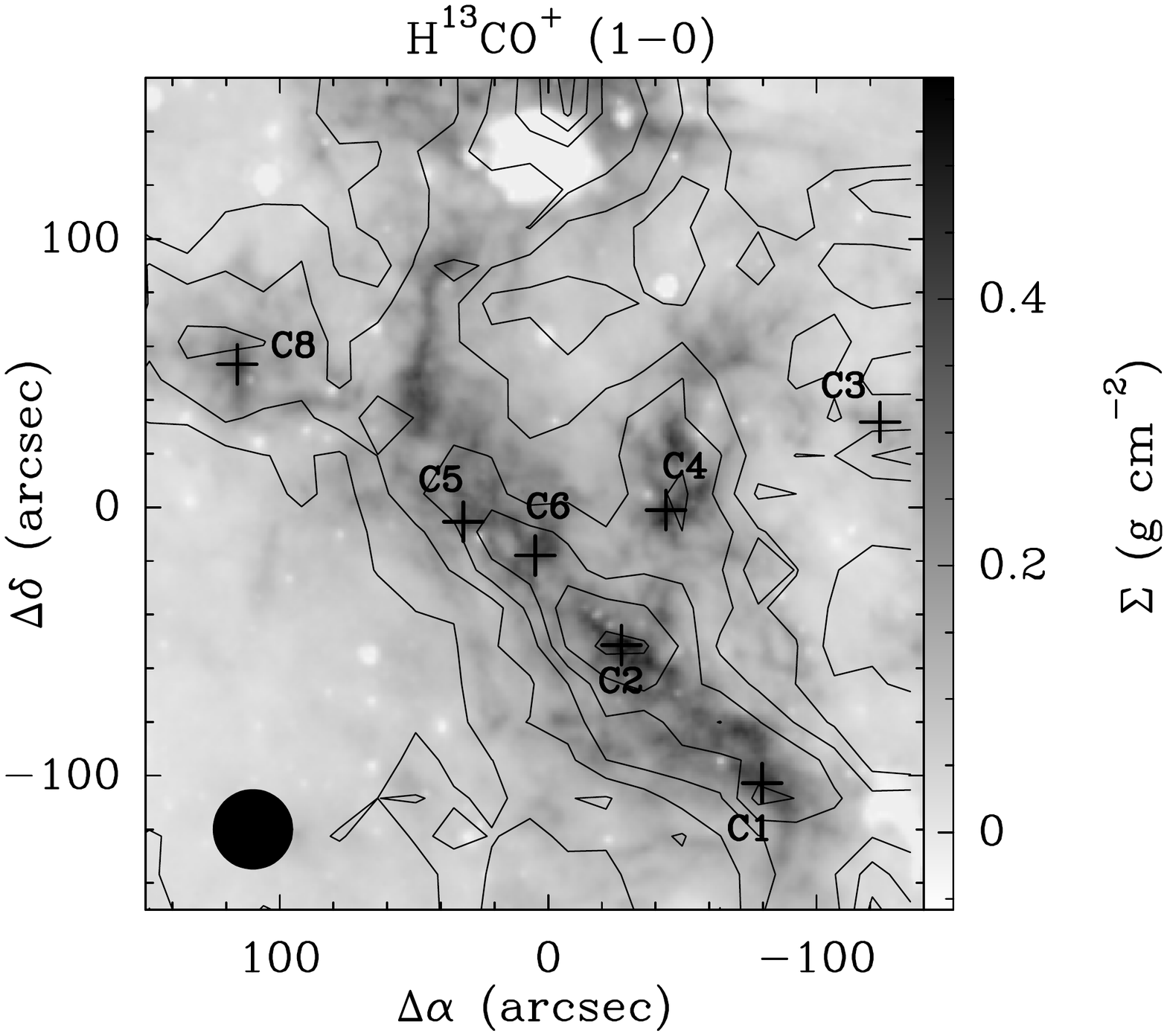}\includegraphics[trim=3cm 2.1cm 4cm 1.7cm,clip=true,scale=0.37]{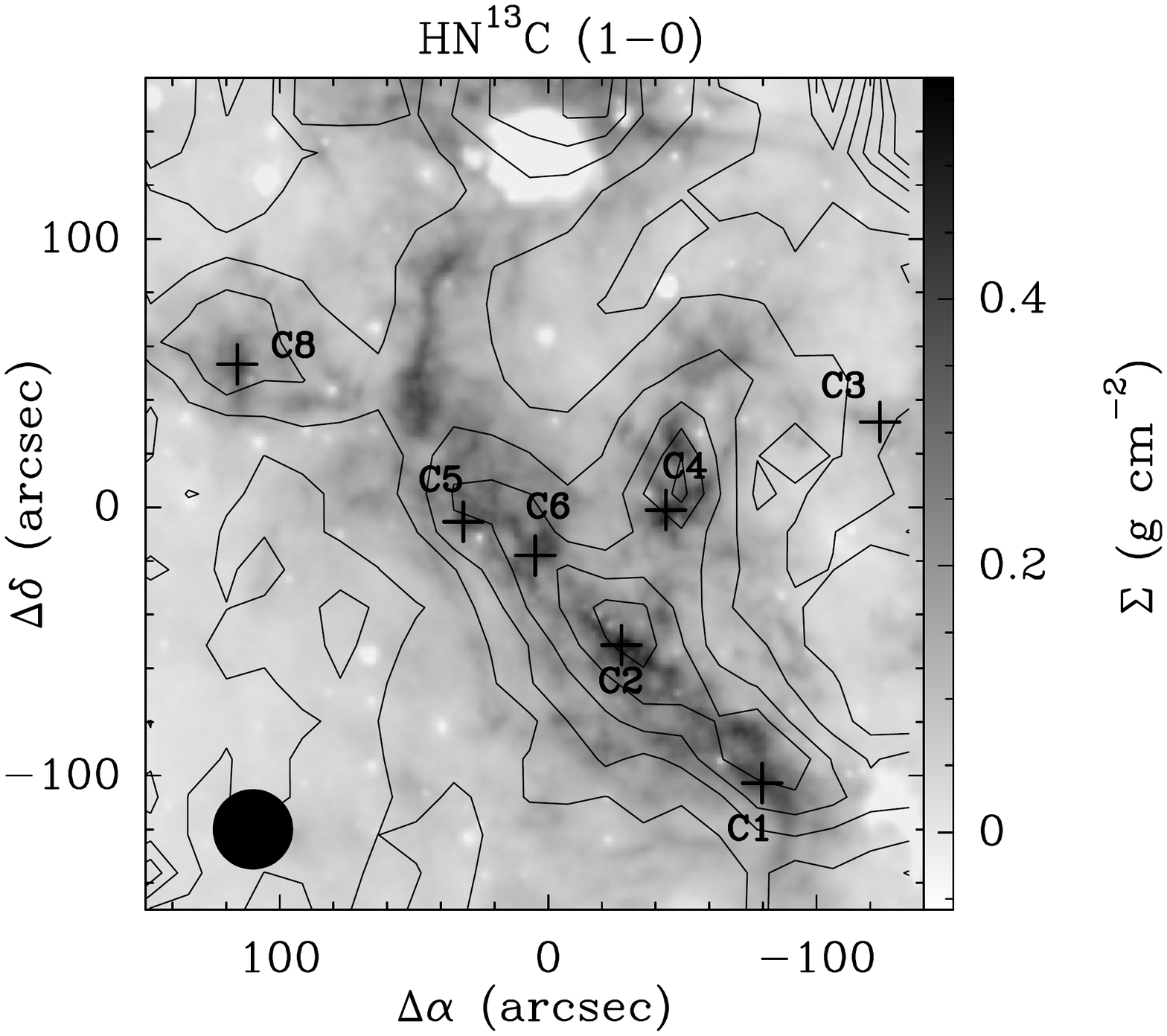}

\includegraphics[trim=-1.5cm 3.1cm -.5cm 4.2cm,clip=true,scale=0.37]{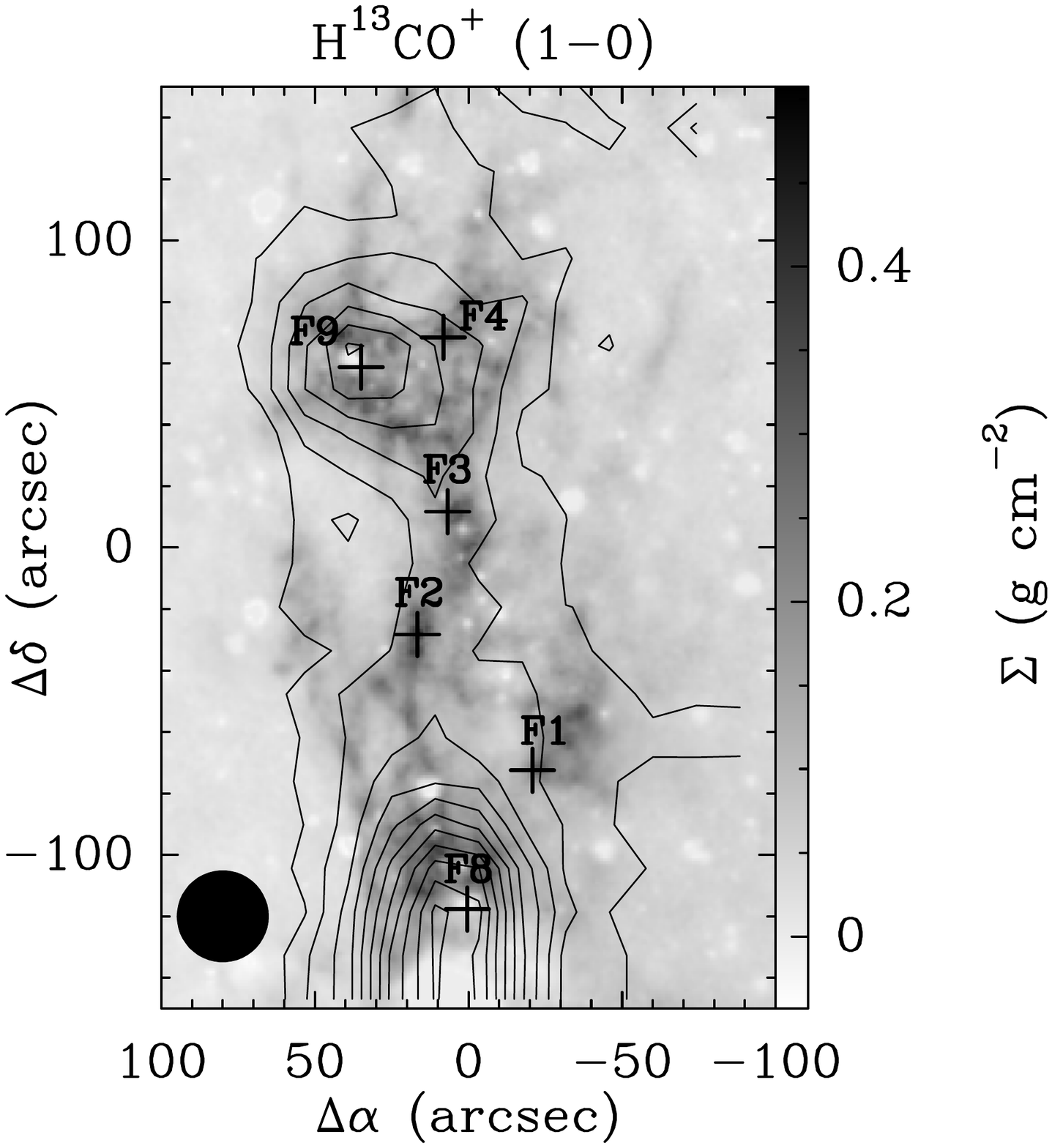}\includegraphics[trim=-1cm 3.1cm 0cm 4.2cm,clip=true,scale=0.37]{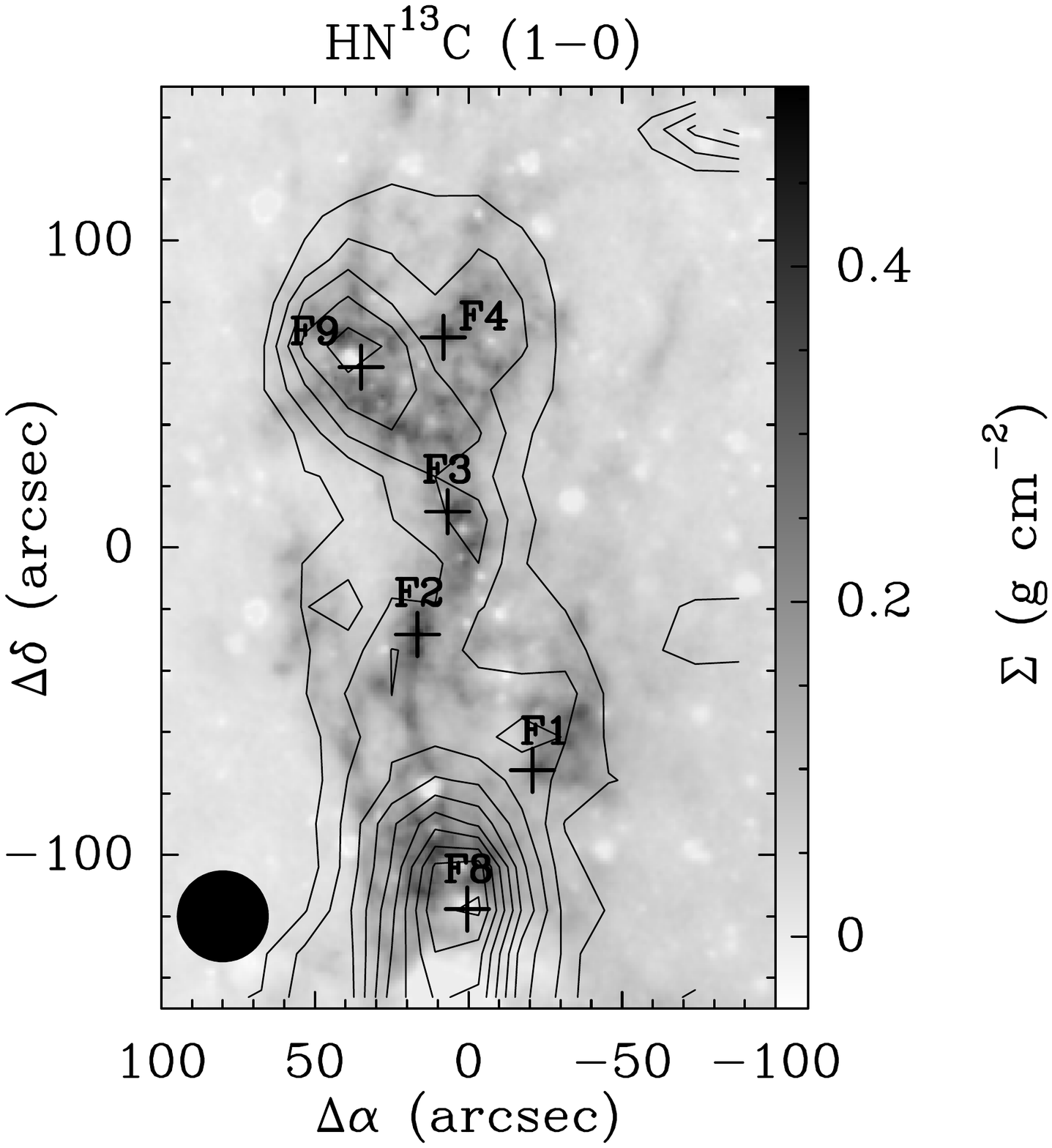}

\includegraphics[trim=1cm 1.1cm 3cm 1.7cm,clip=true,scale=0.37]{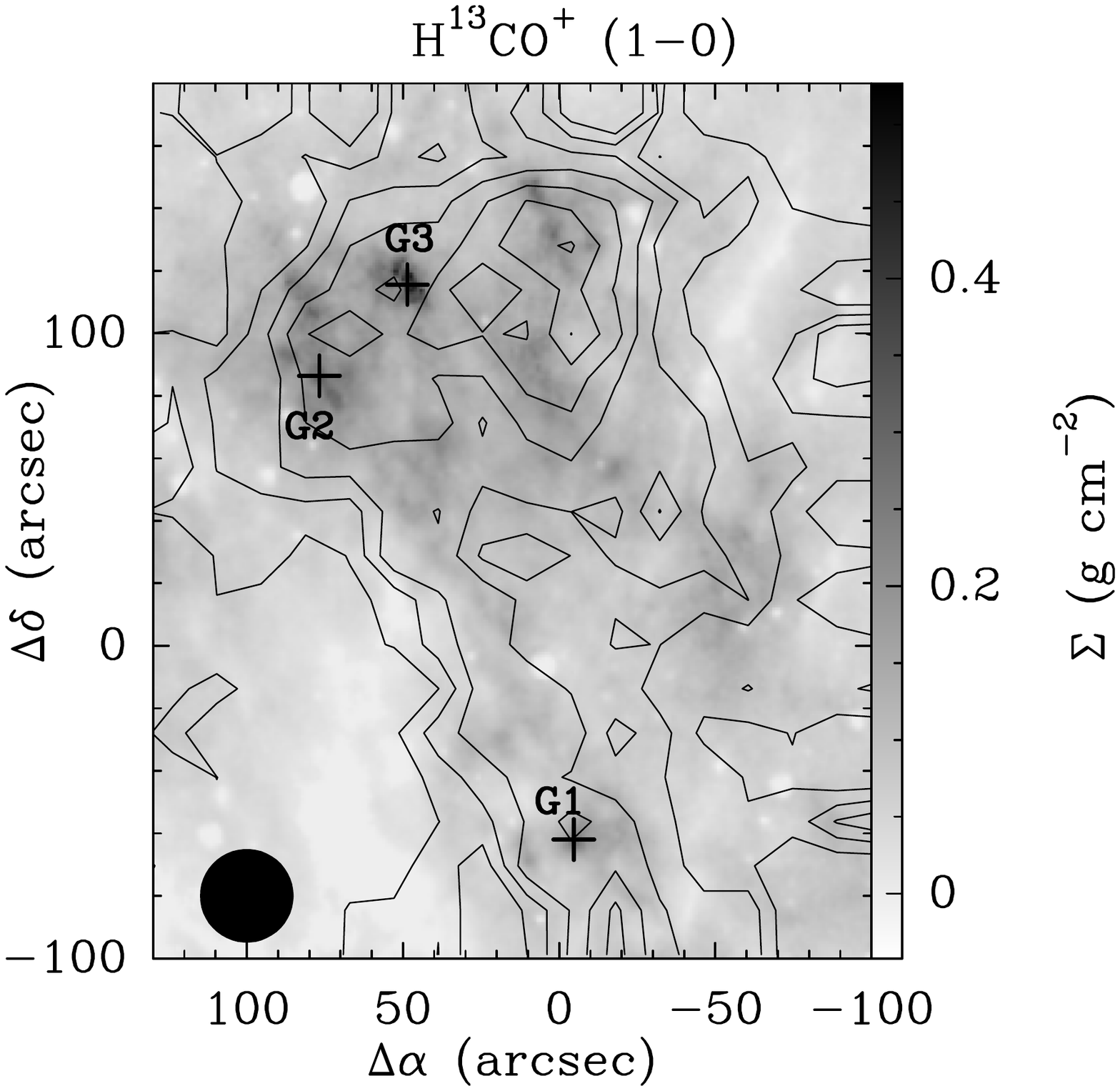}\includegraphics[trim=1cm 1.1cm 1cm 1.7cm,clip=true,scale=0.37]{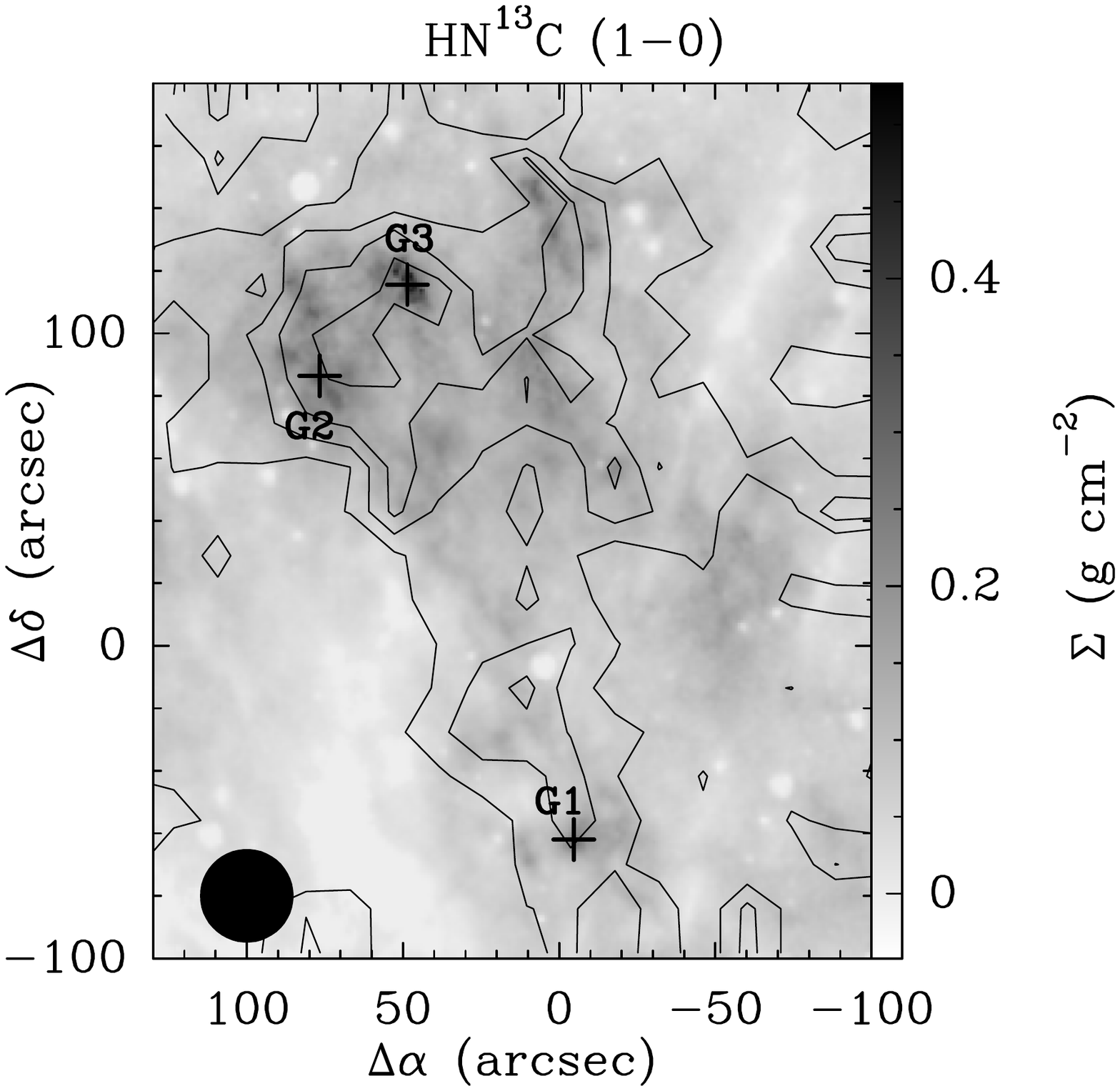}
\caption{Integrated intensity maps of the $\hion$(\onetzeroj) (left) and $\hmol$(\onetzeroj) (right) lines toward cloud C (top panels), F (middle panels) and G (bottom panels). Emission levels (black contours), from 3$\sigma$ by 3$\sigma$, are superimposed on the H$_2$ mass surface density maps (gray scale) obtained by Kainulainen \& Tan (2013). \textit{Cloud C:} The integration range is 76 - 83 km s$^{-1}$ and $\sigma$= 0.10 K km s$^{-1}$ for both molecules. \textit{Cloud F:} Integration range is 51 - 64 km s$^{-1}$ for both tracers and $\sigma$ = 0.11 and 0.15 K km s$^{-1}$ for $\hion$ and $\hmol$ respectively. \textit{Cloud G:} Integration range is 39 - 45 km s$^{-1}$; $\sigma$ = 0.09 and 0.06 K km s$^{-1}$ for $\hion$ and $\hmol$, respectively. The core positions \citep[black crosses;][]{butler2009,butler2012} and the beam sizes (black circles) are shown in all panels.}
\label{Fig1}
\end{figure*}

In Figure~\ref{Fig1}, we report the integrated intensity maps of the $\hion$ (left panels) and $\hmol$ (right panels) emission for clouds C (top), F (middle) and G (bottom). The emission levels (black contours) are superimposed on the H$_2$ mass surface density maps (in gray scale) obtained by \cite{kainulainen2013}.
In all the following maps, the names \citep{butler2009, butler2012} and positions (black crosses; Butler \& Tan 2009 for F8 and F9; Butler \& Tan 2012 for all the other cores) are indicated for all cores.
In all clouds of our sample, the dense gas tracer emission highlights the filamentary morphology of the clouds, closely following the structure detected in extinction (or defined by the H$_2$ mass surface density map). For clouds C (integrated velocity range 76 - 83 km s$^{-1}$) and F (velocity range 51 - 64 km s$^{-1}$), the $\hion$ and $\hmol$ emission shows similar spatial distributions, with peaks mainly associated with the cores present in the clouds. 

In contrast to clouds C and F, cloud G (integrated velocity range 39 - 45 km s$^{-1}$) shows a slightly different spatial distribution for $\hion$ (left) and $\hmol$ (right). The $\hmol$ emission peaks coincide with the location of the two north-eastern cores G2 and G3, while the $\hion$ emission peaks toward the north-west of the cloud at offset (0\arcsec, 130\arcsec). The distance between the $\hmol$ and the $\hion$ emission peaks is 54\arcsec \space (0.76 pc), larger than the angular resolution of our maps (30\arcsec) and enough to conclude that the $\hion$ emission peak is not associated with the two north-eastern cores of the cloud. To the best of our knowledge, no cores are present at the position of the $\hion$ emission peak.

\subsubsection{Sub-Structures identification}\label{dense_gas_filaments}
\begin{figure*}
\centering
\includegraphics[trim= 1cm 8cm 1cm 8cm,clip=true,scale =1.]{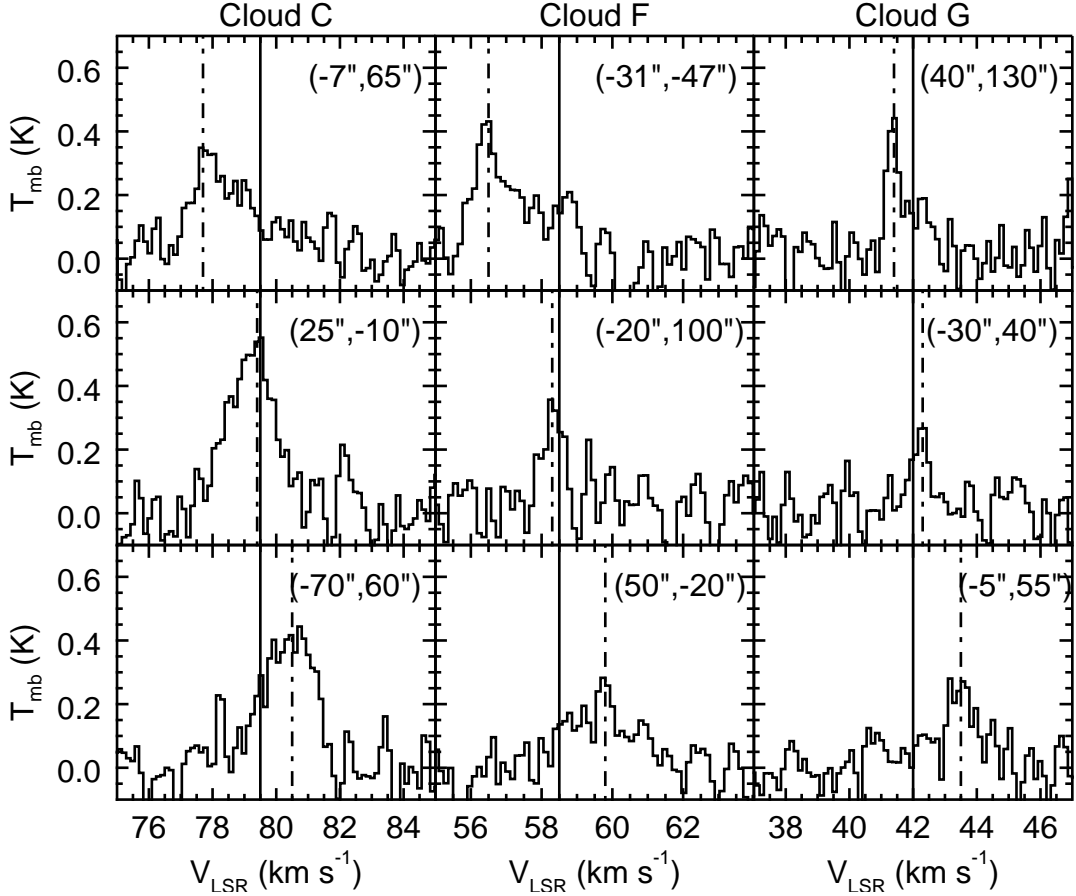}\\
\caption{$\hmol$ spectra for clouds C (left panel), F  (middle panel) and G (right panel) extracted toward several positions and highlighting the presence of multiple velocity components. The vertical solid lines correspond to the central velocity of the clouds while the vertical dot-dashed lines indicate the central velocity of the spectrum peaks. The same result has been found for the $\hion$ emission in all the three clouds and are not shown for simplicity.}
\label{fig2}
\end{figure*}

Visual inspection of the $\hion$ and $\hmol$ spectra throughout each IRDC confirms that a single Gaussian component is a poor approximation to the spectral line profiles observed at certain locations. Each cloud appears to exhibit a dominant velocity component which largely follows the morphology of the IRDC as seen in extinction. For clouds C, F, and G these are situated at $\sim 79.5$ km s$^{-1}$, $\sim 58.5$ km s$^{-1}$, and $\sim 41.9$ km s$^{-1}$, respectively. However, each cloud shows additional, morphologically-distinct features which are shifted from the dominant component in either position or velocity (or both). These additional features lead to asymmetries in the line profiles of the $\hmol$ and $\hion$ lines. In Figure~\ref{fig2}, $\hmol$ spectra extracted toward different positions across the clouds are displayed. Form this Figure, it is clear that the spectra exhibit different central velocities toward different region in the clouds, indicating the presence of multiple velocity structures. These components show differences in the centroid radial velocities \hbox{$>$ 1 km s$^{-1}$}, i.e. more than 2/3 the mean emission linewidth in all clouds and more than 7$\times$ and 3.5$\times$ the velocity resolution ($\delta v$) for cloud F, and clouds C and G, respectively.

\begin{figure*}
\centering
\includegraphics[angle=0,scale=0.8,trim= 0cm 3cm 0cm 3cm, clip=True]{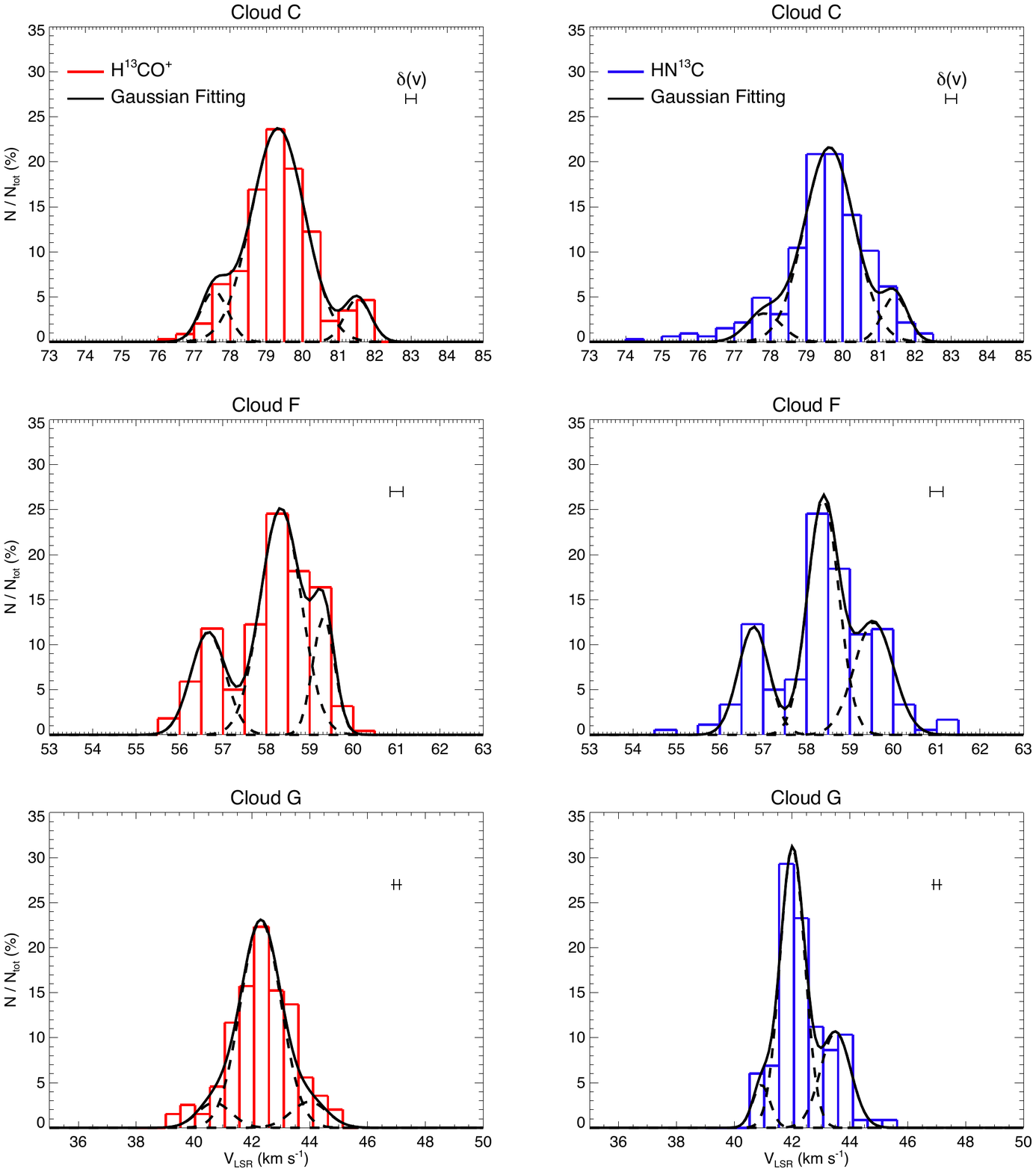}
\caption{Velocity distributions of the $\hion$ (red) and $\hmol$ (blue) emissions as obtained for cloud C (top panel), cloud F (middle panel) and cloud G (bottom panel). The histograms show the percentage of emission lines having V$_{LSR}$ falling within the $\delta$V of the bin. Bin sizes are 0.5 km s$^{-1}$ for the three clouds, which corresponds to 1/3 of mean intensity-weighted linewidth for the two tracers. In all panels the mean uncertainty in the central velocities of the fittings is indicated.} 
\label{fig3}
\end{figure*}

In order to determine the spatial extent and kinematics of these sub-structures, we further investigate the overall distribution of the central radial velocities and line widths of $\hion$ and $\hmol$ extracted using SCOUSE across the clouds. In Figure~\ref{fig3}, we show histograms of the velocity distributions obtained for $\hion$ (left panels, in red) and $\hmol$ (right panels, in blue) toward clouds C (top), F (middle) and G (bottom). 
The bin size in the x-axis corresponds to 1/3 of the mean intensity-weighted linewidth, obtained from the SCOUSE results (see Section 4.1.2 and Table~\ref{tab4}. This corresponds to 0.5 km s$^{-1}$ for the three clouds. In the y-axis we plot the percentage of emission line components having central velocities falling within the bin. For each histogram, we also report the mean uncertainty on the central velocity, obtained by averaging the central velocity uncertainties derived by SCOUSE for each fitted line component.

Asymmetries in the distribution of the radial velocities for each cloud are more pronounced in Figure~\ref{fig3}. The velocity distribution of $\hion$ and $\hmol$ are best fitted with three Gaussians (black lines). Using $\hmol$ (right panels), where the asymmetries are most prominent, we define a velocity range for each cloud component as follows: 

\begin{equation}
\Delta V = V^{gf}_{LSR} \pm (1.5 \times \sigma^{gf})
\label{eq1}
\end{equation}

\noindent
where $\Delta V$ is the velocity range of the sub-structures, $V^{gf}_{LSR}$ is the central velocity and $\sigma^{gf}$ the standard deviation of the fitted Gaussian (\textit{gf} = gaussian fitting). 
In Table~\ref{tab3}, we report the velocity ranges defined for the sub-structures and their mean velocities obtained as (Jim\'enez-Serra et al. 2014): 

\begin{equation}
<V> = \frac{\Sigma_{i} V_{i} I_{i}}{\Sigma_{i} I_{i},} 
\label{eq2}
\end{equation}

\noindent 
The goodness of fit is provided by calculating the mean of the main residual values defined as:

\begin{equation}
Mean \quad Residual = \frac{\Sigma |G_i - (N/N_{tot})_i|}{N_{bin}}
\label{eq2}
\end{equation}

\noindent
where G$_i$ indicates the value of the multi-gaussian fitting at the i$^{th}$ bin, (N/N$_{tot}$)$_i$ is the number of fitted lines falling within the i$^{th}$ bin and N$_{bin}$ is the number of bins of the histogram.

Since both $\hion$ and $\hmol$ are predominantly tracers of high-density material, and each component is well separated in velocity (Figure~\ref{fig3}), the chances of confusion and/or blending between components is small. 
However, we note that a more comprehensive description of the gas kinematics of each cloud using more abundant, lower density gas tracers, and more sophisticated techniques \citep[see][]{henshaw2014}, will be presented in two forthcoming papers (Barnes et al., in prep.; Henshaw et al., in prep.). 

\begin{table*}
\centering
\caption{Mean intensity-weighted radial velocities and velocity ranges of the identified sub-structures in clouds C, F and G. The mean residual of the multi-Gaussian fitting is reported as indication of fitting goodness.}
\begin{tabular}{cccccccc}
\hline
\hline
 Cloud & $V_{1}$ & $\Delta V_{1}$ & $V_{2}$& $\Delta V_{2}$ & $V_{3}$ & $\Delta V_{3}$ & Mean Residual\\
 & (km s$^{-1}$) & (km s$^{-1}$)& (km s$^{-1}$)& (km s$^{-1}$)&(km s$^{-1}$)& (km s$^{-1}$) &\\
\hline
C & 77.8  & 77-78.5  & 79.5  & 79-80    & 81.4 &81-82 & 0.30\\
F & 56.6  & 56-57    & 58.5  & 58-59    & 59.5 &59-60 & 0.13\\
G & 41.0  & 40.5-41.5& 41.9  & 41.5-42.5& 43.5 &43-44 & 0.13\\
\hline
\end{tabular}
\label{tab3}
\end{table*}

\begin{figure*}
\centering
\includegraphics[angle=0,scale=0.32,trim=2.7cm 0cm 6cm 1cm,clip=true]{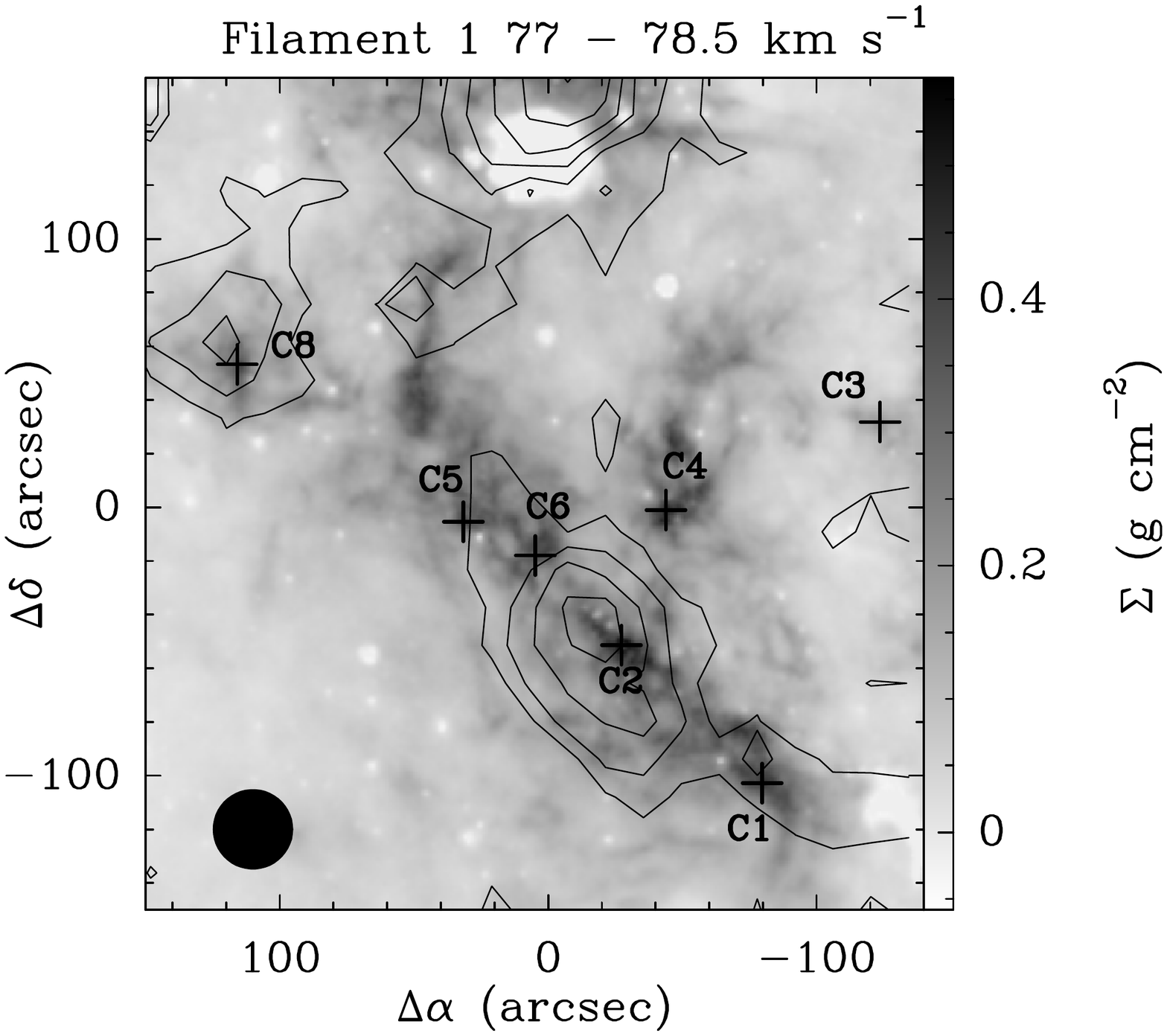}\includegraphics[angle=0,scale=0.32,trim=3cm 0cm 6cm 1cm,clip=true]{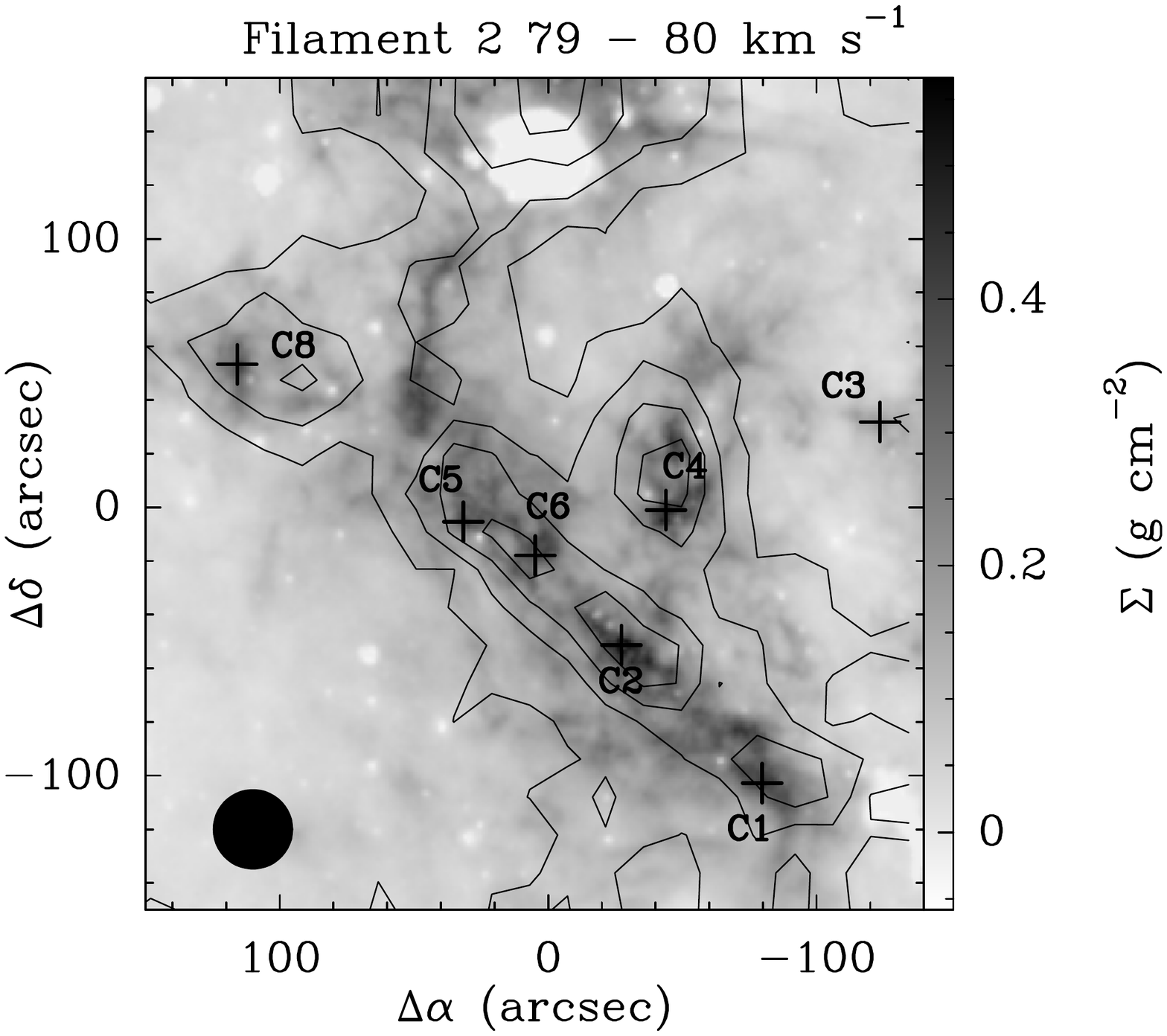}\includegraphics[angle=0,scale=0.32,trim=3cm 0cm 1cm 1cm,clip=true]{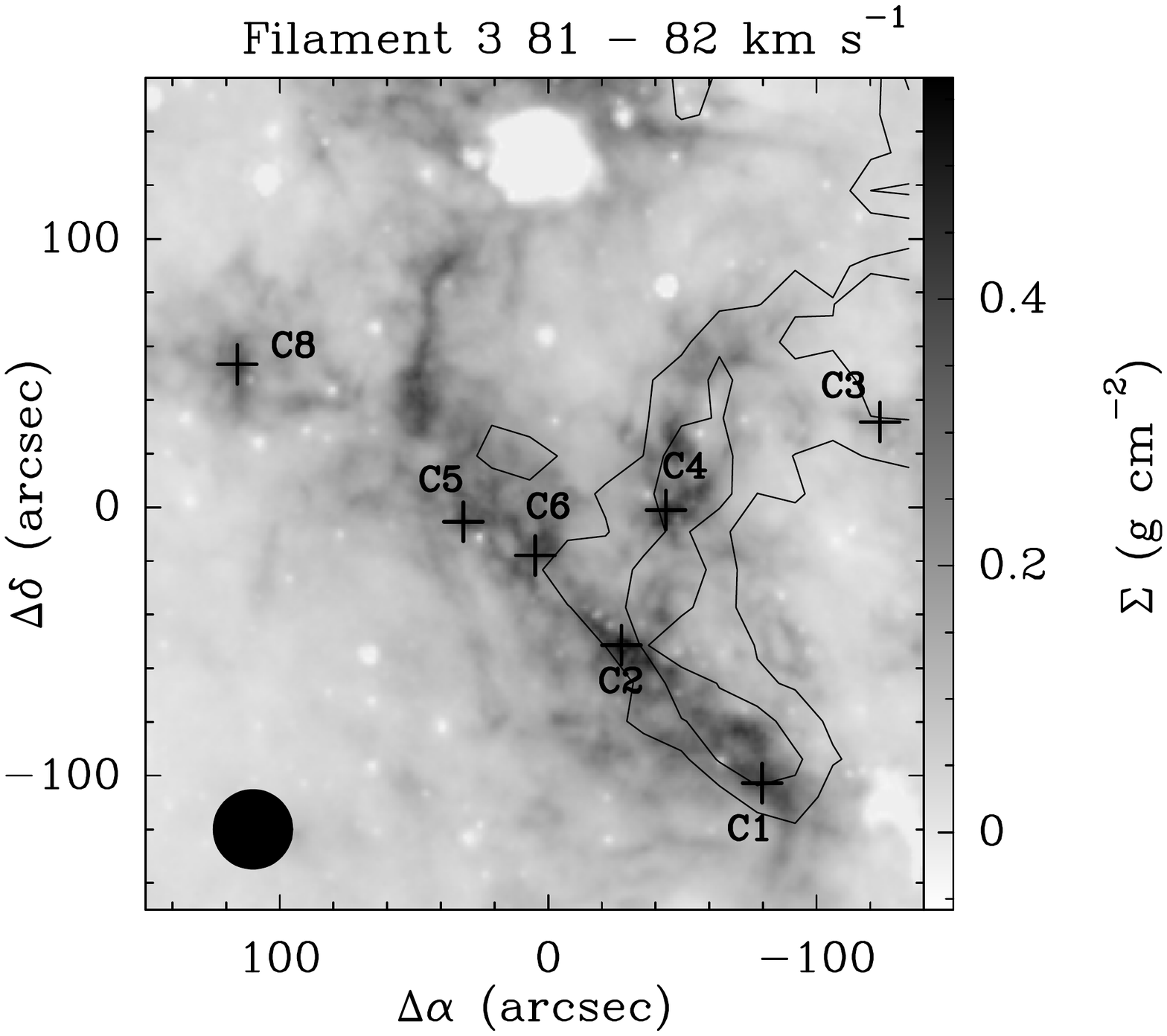}\\

\includegraphics[angle=0,scale=0.32,trim=0cm 0cm 4cm 3cm,clip=true]{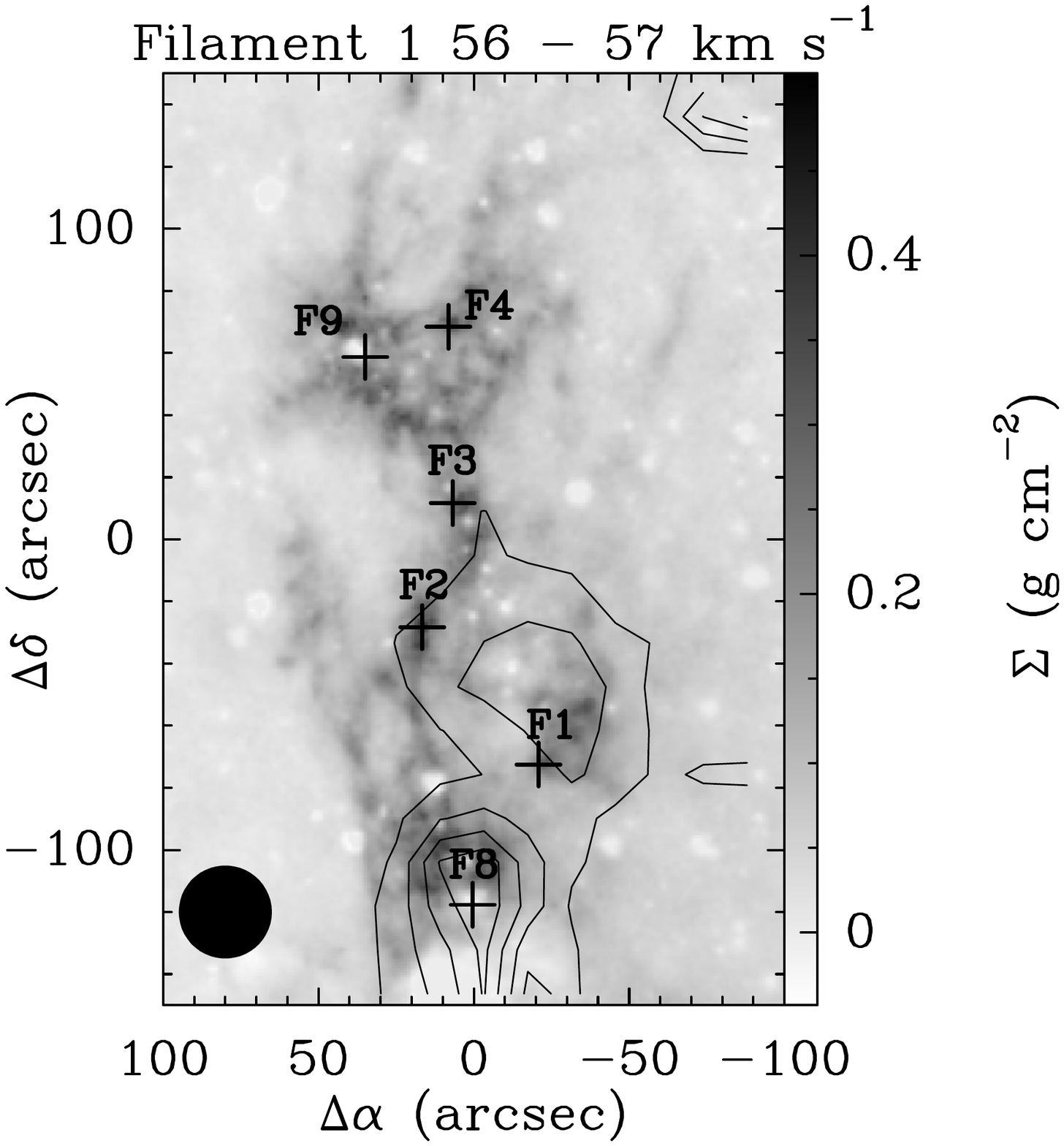}\includegraphics[angle=0,scale=0.32,trim=-1cm 0cm 4cm 3cm,clip=true]{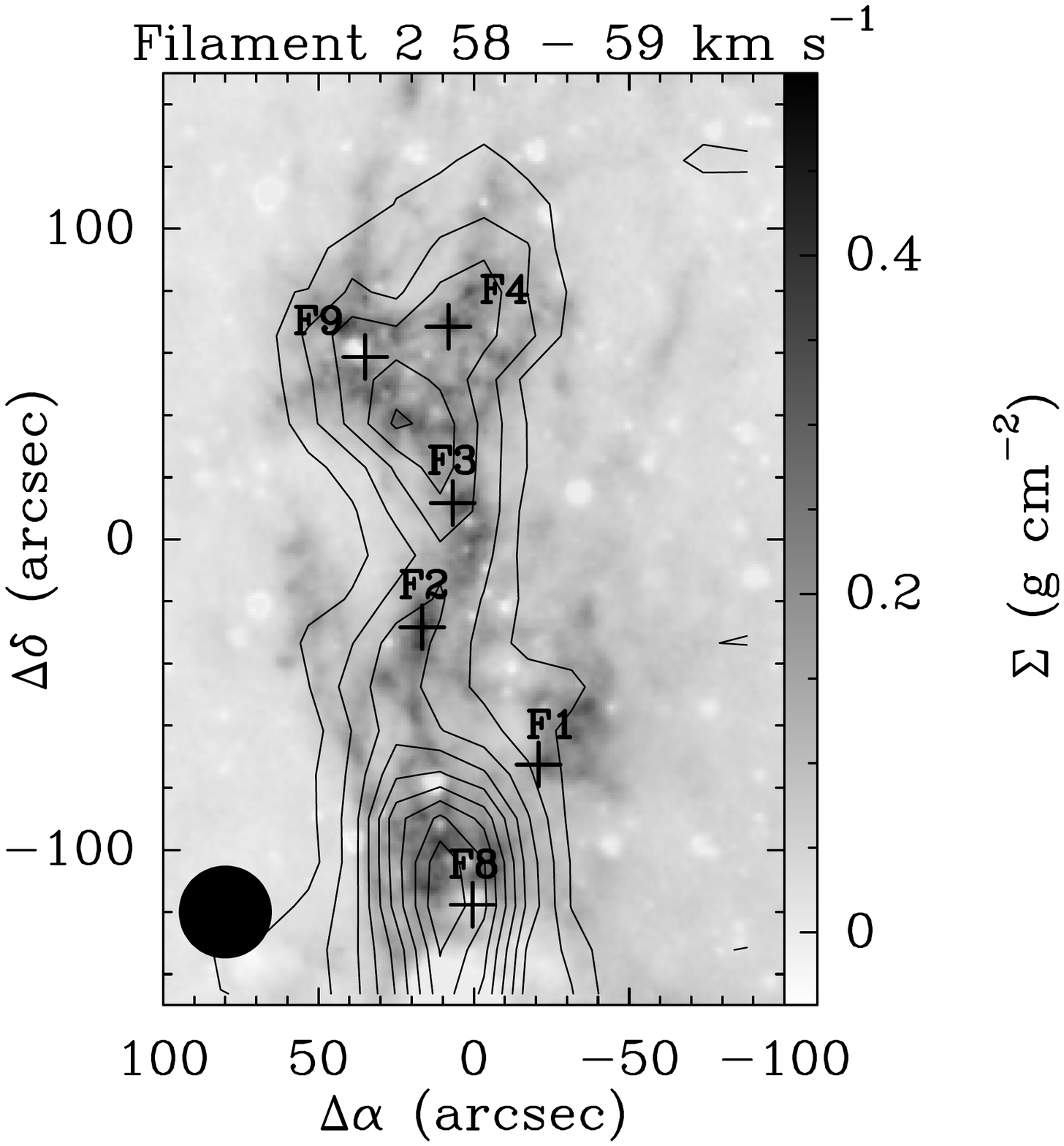}\includegraphics[angle=0,scale=0.32,trim=-1cm 0cm 0cm 3cm,clip=true]{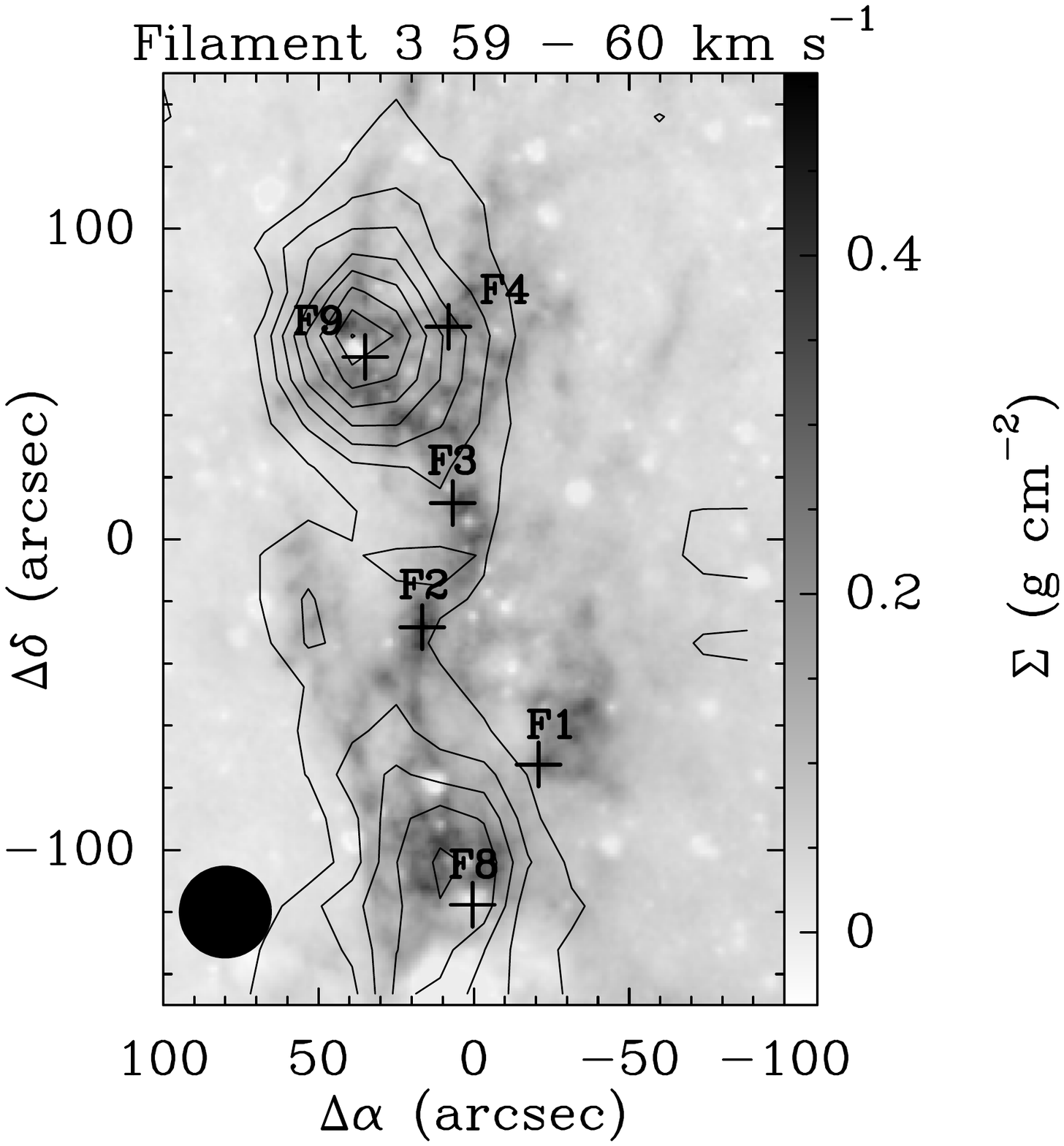}\\

\includegraphics[angle=0,scale=0.32,trim=2cm 0cm 7cm 1cm,clip=true]{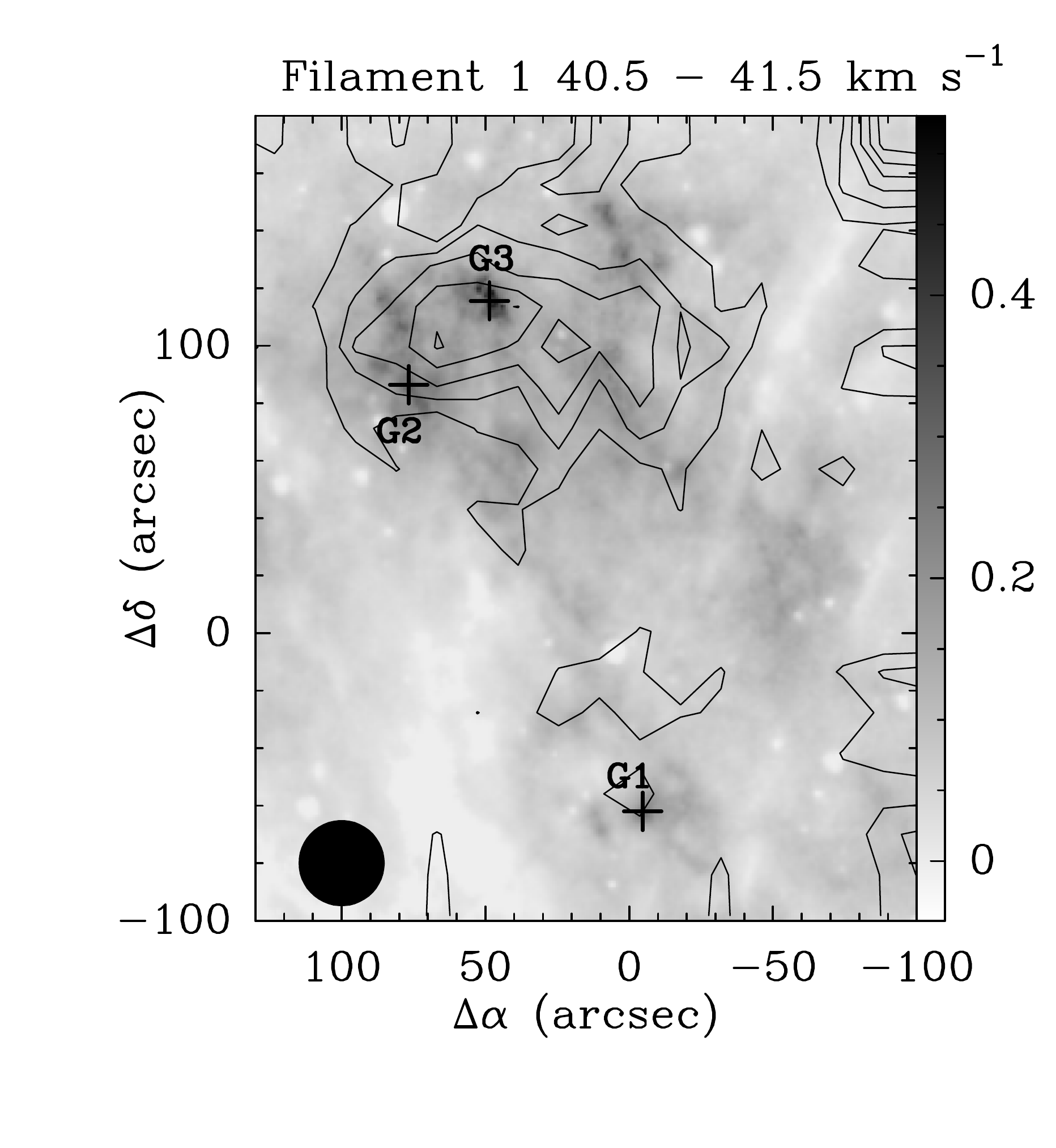}\includegraphics[angle=0,scale=0.32,trim=3cm 0cm 7cm 1cm,clip=true]{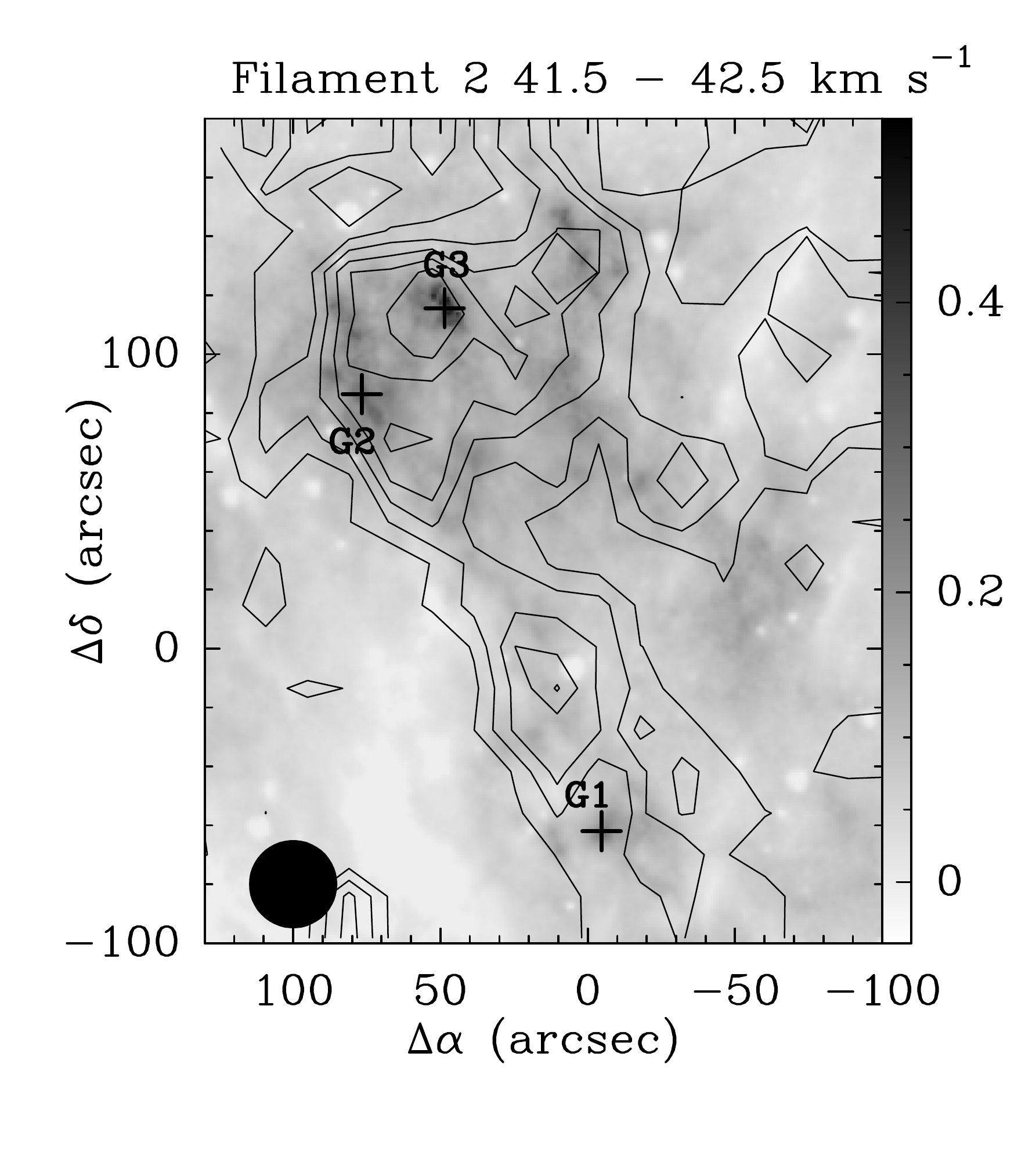}\includegraphics[angle=0,scale=0.32,trim=3cm 0cm 0cm 1cm,clip=true]{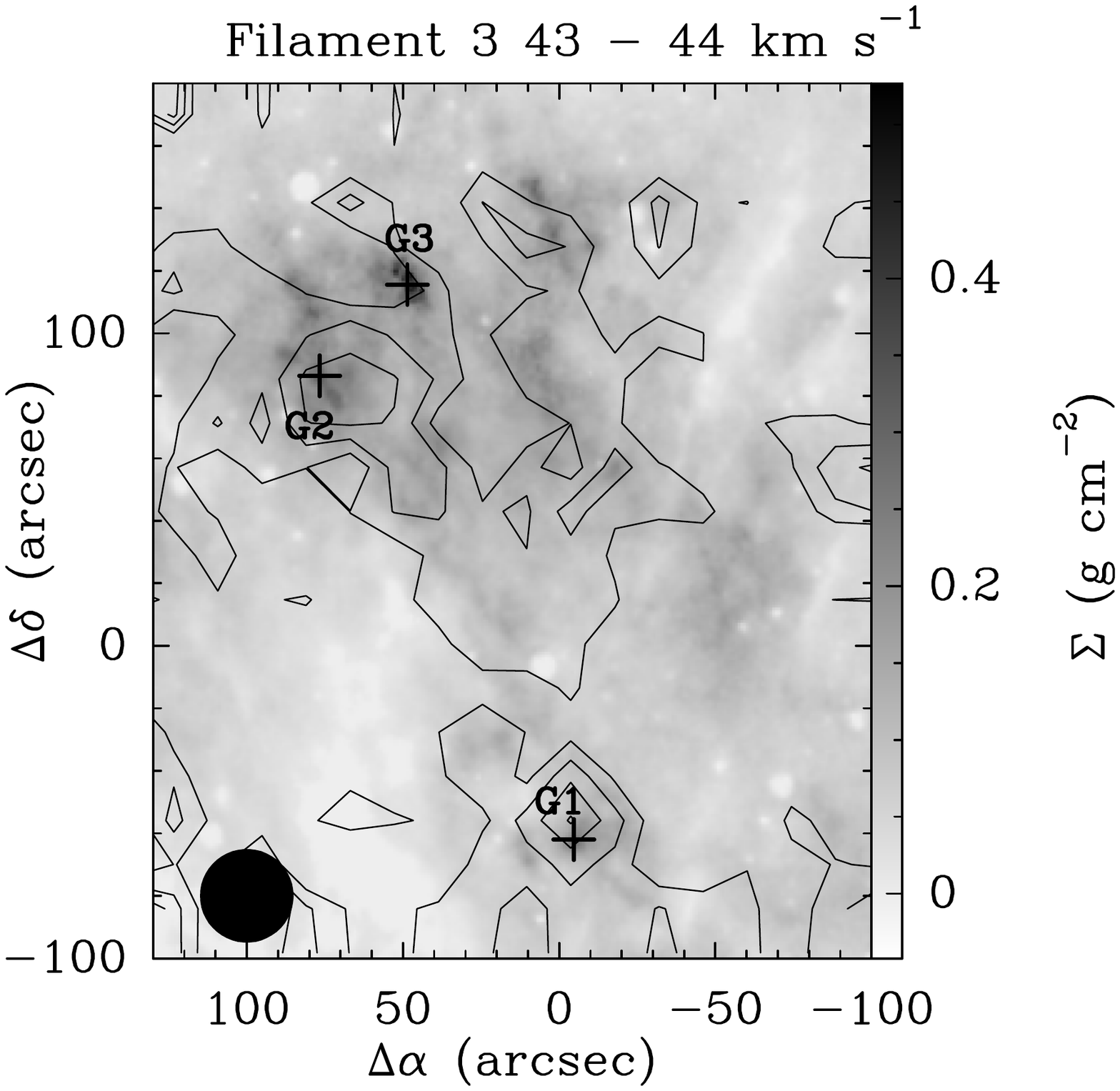}\\

\caption{ Integrated intensity maps of the $\hmol$ emission in clouds C (top panels), F (middle panels) and G (bottom panels) for the identified velocity sub-structures. The emission levels (black contours) are superimposed on the H$_2$ mass surface density map of Kainulainen \& Tan (2013). The integrated velocity ranges are indicated in every panel and the emission levels go from 3$\sigma$ by 3$\sigma$, where $\sigma$ = 0.04 K km s$^{-1}$ for cloud C, $\sigma$ = 0.04 K km s$^{-1}$ for cloud F and $\sigma$ = 0.02 K km s$^{-1}$ for cloud G. The beam sizes are shown as black circles in all panels.}
\label{fig4}
\end{figure*}

These sub-structures can be visualized in Figure~\ref{fig4} where we show the integrated intensity maps obtained for $\hmol$ for the velocity ranges given in Table~\ref{tab3}. 
The identified sub-structures have filamentary shapes and change in position as a function of velocity with respect to the IRDC seen in extinction. In particular, for cloud C (uppermost panels) the central and brightest sub-structure at 79.5 km s$^{-1}$ extends from south-west to north, following the structure of the cloud. The blue-shifted sub-structure with mean velocity 77.8 km s$^{-1}$ (uppermost left panel), however, is fainter and its emission arises mainly from the eastern part of the cloud peaking at cores C5, C2 and C1. 
The red-shifted sub-structure peaking at 81.4 km s$^{-1}$ (uppermost right panel) shifts toward the western part of the cloud and runs from the south-west to the north-west exhibiting an arch-like shape. Its emission seems to be associated with cores C4 and C1. The three sub-structures spatially overlap toward the central chain of cores. A similar behavior was also noted for IRDC G035.39-00.33 (or cloud H) by \cite{jimenezserra2014}.  
For cloud F (Figure~\ref{fig4}, middle panels), the blue-shifted sub-structures with mean velocity 56.6 km s$^{-1}$ (left panel) is the faintest, least extended, and arises mainly from the western part of the cloud. As we move to more red-shifted velocities, the central elongated sub-structure at 58.5 km s$^{-1}$ becomes apparent (middle panel), showing the brightest emission and following the filamentary structure of the cloud seen in the H$_2$ mass surface density map. The red-shifted sub-structure peaking at 59.5 km s$^{-1}$ (right panel) is also elongated but the bulk of the emission shifts towards the east of the cloud. 

Very peculiar is the spatial distribution of the sub-structures in cloud G (bottom panels in Figure~\ref{fig4}). The blue-shifted sub-structures centered at 41.0 km s$^{-1}$ (left panel) is the least extended and its emission clearly originates from the G3 core. This is also true for the central sub-structure at 41.9 km s $^{-1}$ (middle panel), although it shows two extensions toward the south-west following the H$_2$ mass surface density map. 

In contrast, the red-shifted sub-structure of cloud G at 43.5 km s$^{-1}$ (right panel) shows a U-shape facing north-east unseen in the other sub-structures. 

\subsubsection{Line Width Distribution}\label{dense_gas_linewidth}
\begin{figure*}
\centering
\includegraphics[trim= 0.2cm 6cm 0cm 5cm,clip=true,scale=.8]{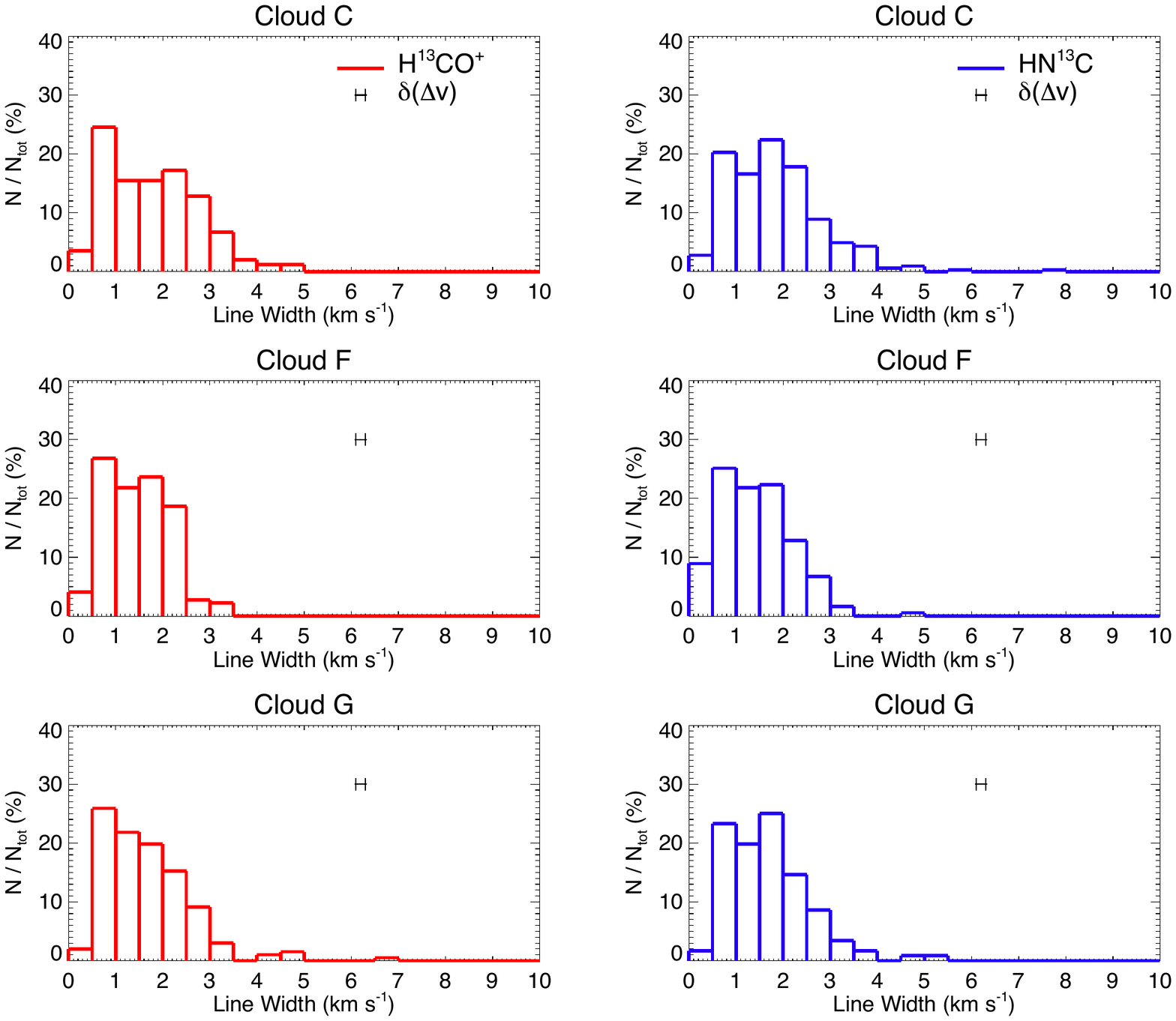}
\caption{Line width distributions of the $\hion$ (left panels in red) and $\hmol$ (right panels in blue) emission measured for clouds C (top panels), F (middle panels) and G (bottom panels). The histograms show the percentage of emission lines having linewidths falling within each bin. Bin sizes are 0.5 km s$^{-1}$ for the three clouds, corresponding to 1/3 of the mean intensity-weighted linewidth. In each panel the mean uncertainty in the linewidth of the single fittings is indicated.}
\label{fig5}
\end{figure*}

We now investigate whether the velocity components of the dense gas show any difference in their measured line widths. In Figure~\ref{fig5}, we report the line width distributions for both $\hmol$ (in blue) and $\hion$ (in red) observed toward clouds C (top panel), F (middle panel) and G (bottom panel). The bin size has been evaluated as in Figure~\ref{fig3} and the percentage of emission lines having line widths falling in every bin is shown in the y-axis. For each histogram, we report the mean  uncertainty in the linewidth, derived averaging the uncertainty in all the linewidths derived by SCOUSE. From Figure~\ref{fig5}, we find a smooth distribution which peaks between 1 and 2 km s$^{-1}$ with no multiple line width components. The line widths for both molecules are in any case narrower than 5 km s$^{-1}$. 

The overall intensity-weighted average line width, and the average linewidth of each sub-structure found in the three clouds, can be calculated as (Jim\'enez-Serra et al. 2014): 

\begin{equation}
<\Delta v> = \frac{\Sigma_{i} \Delta v_{i} I_{i}}{\Sigma_{i} I_{i},}
\label{eq3}
\end{equation}

\noindent
where $\Delta v$ is line width and I$_{i}$ is the intensity measured for each velocity component in each position of the data cube. The obtained values are listed in Table~\ref{tab4}.

From Table~\ref{tab4}, the difference between the average line width of $\hion$ and $\hmol$ is always $< 2\times \delta v$ for the three clouds, which implies that the dynamics of ion species is not significantly different from that of neutral molecules. Therefore, although these differences have been reported in more evolved high-mass star-forming regions \citep[as e.g. DR21; see][]{hezareh2010}, we do not find any evidence for this phenomena in the quiescent gas of IRDCs.  

\subsection{Shock tracer emission: SiO and CH$_3$OH}\label{shock_emission}
\begin{figure*}
\centering
\includegraphics[trim=1cm 2.1cm 2cm 1.5cm,clip=true,scale=0.36]{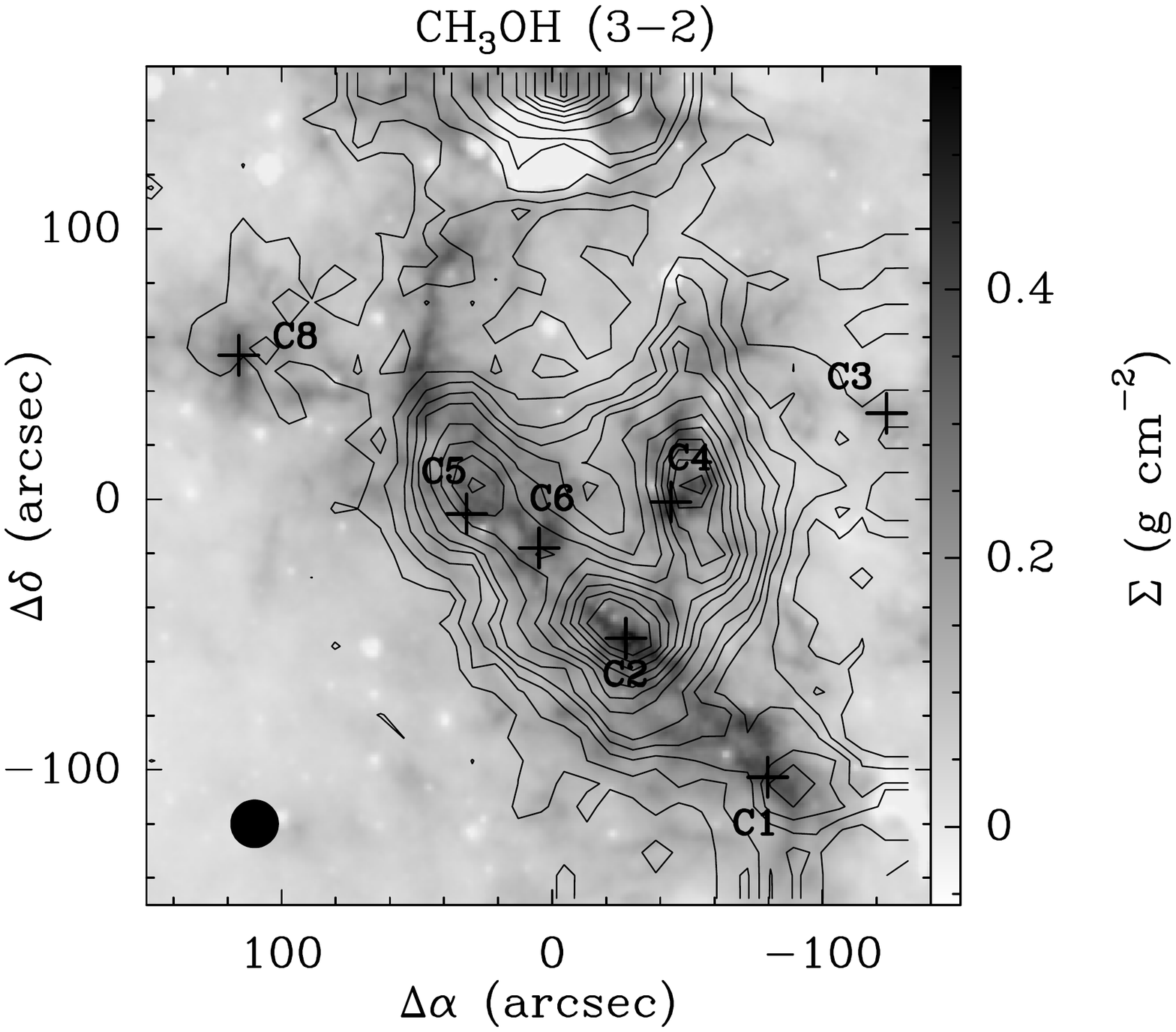}\includegraphics[trim=3cm 2.1cm 2cm 1.5cm,clip=true,scale=0.36]{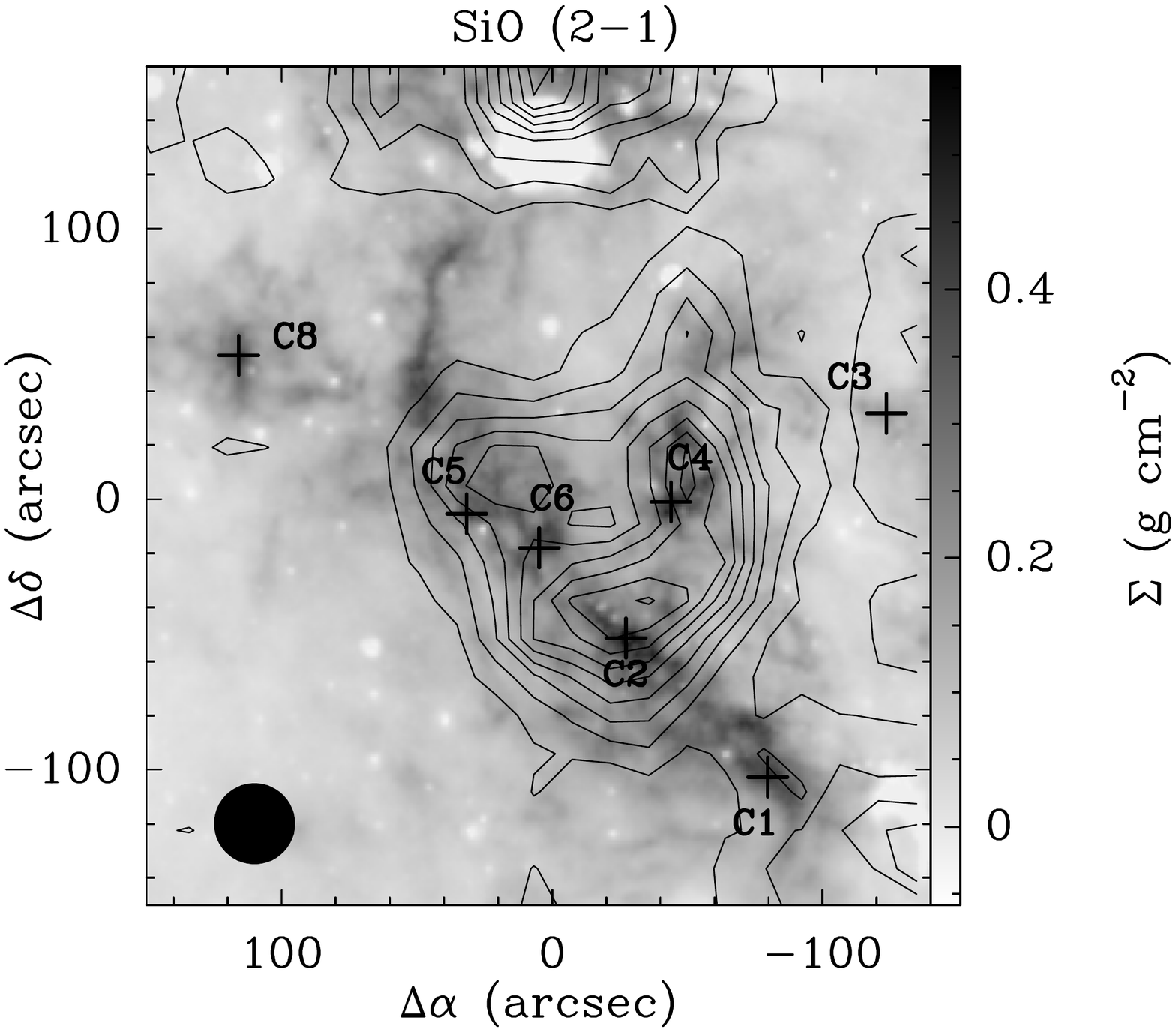}

\includegraphics[trim=0.5cm 3.7cm -3cm 4cm,clip=true,scale=0.36]{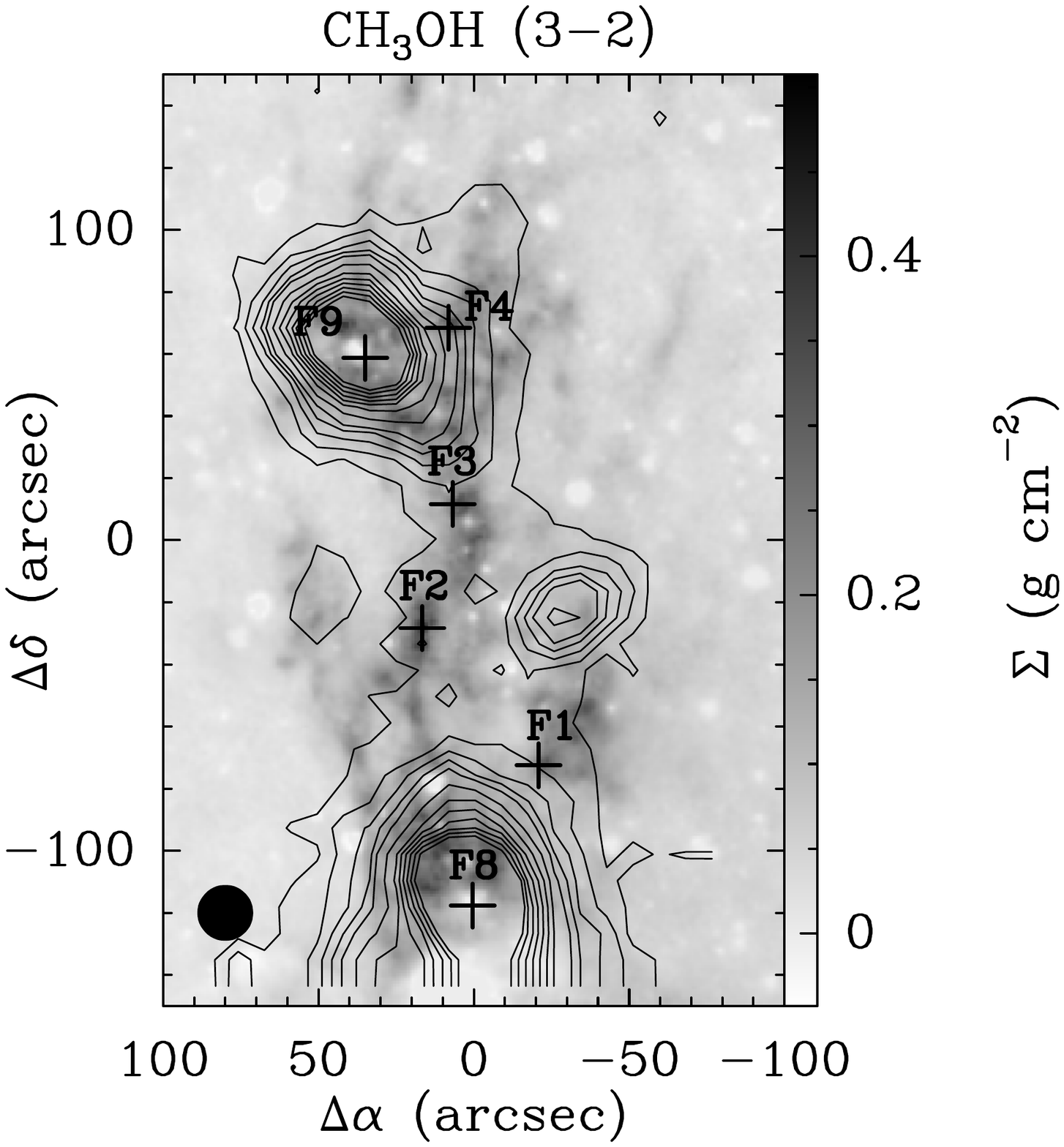}\includegraphics[trim=2cm 3.7cm 0cm 4.cm,clip=true,scale=0.36]{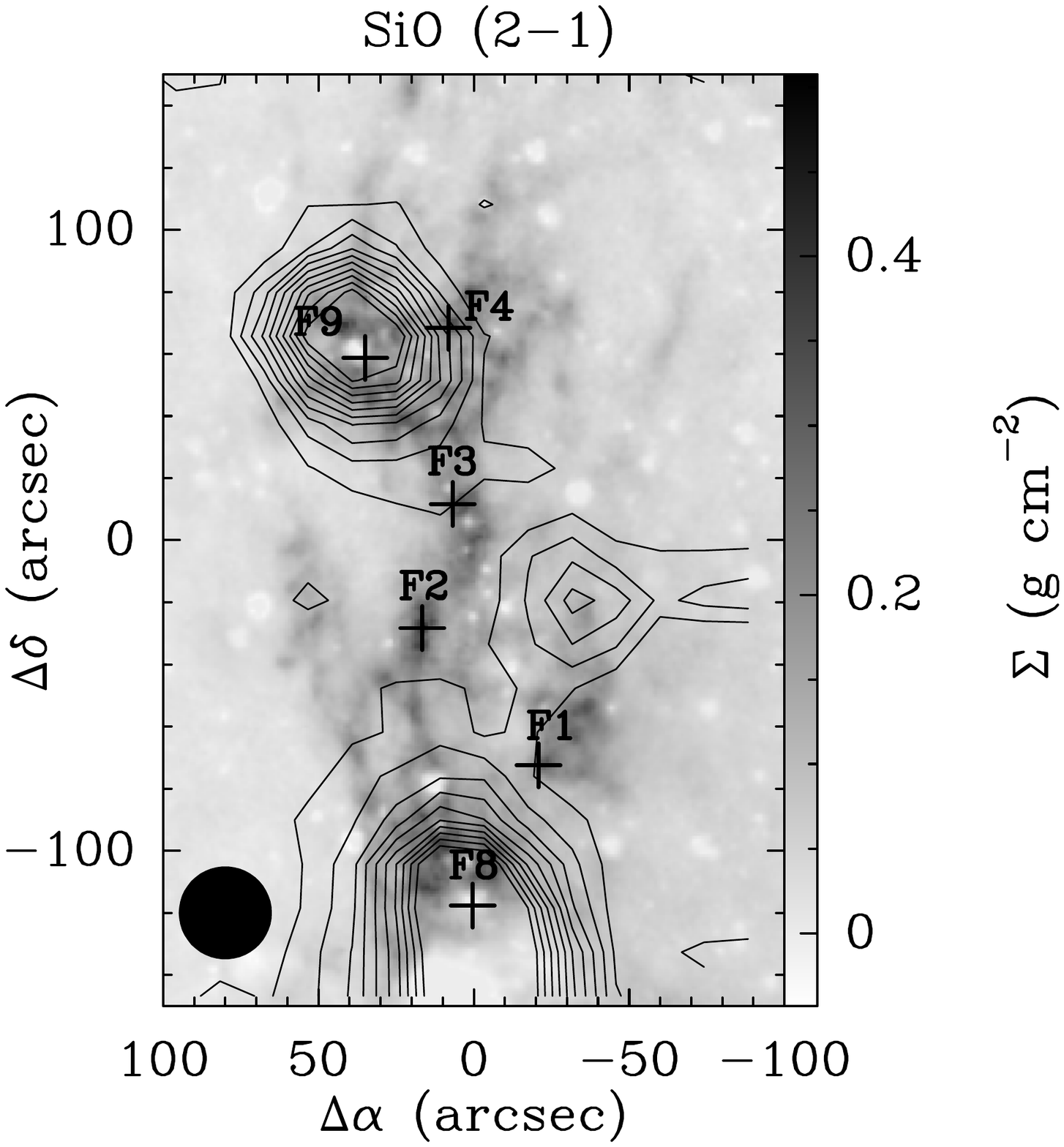}

\includegraphics[trim=1cm 2cm 3cm 1.2cm,clip=true,scale=0.36]{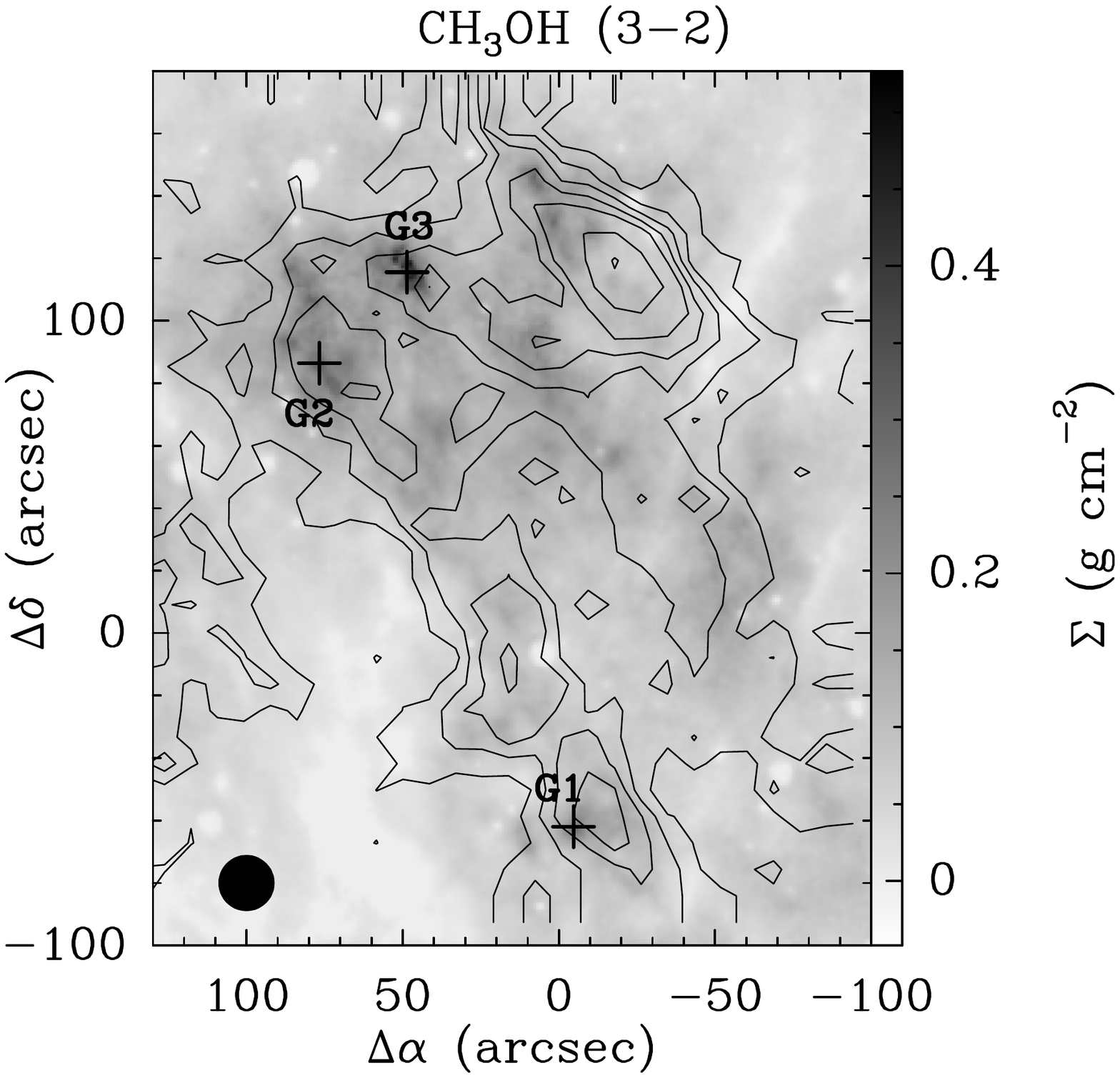}\includegraphics[trim=1.9cm 2cm 1cm 1.2cm,clip=true,scale=0.36]{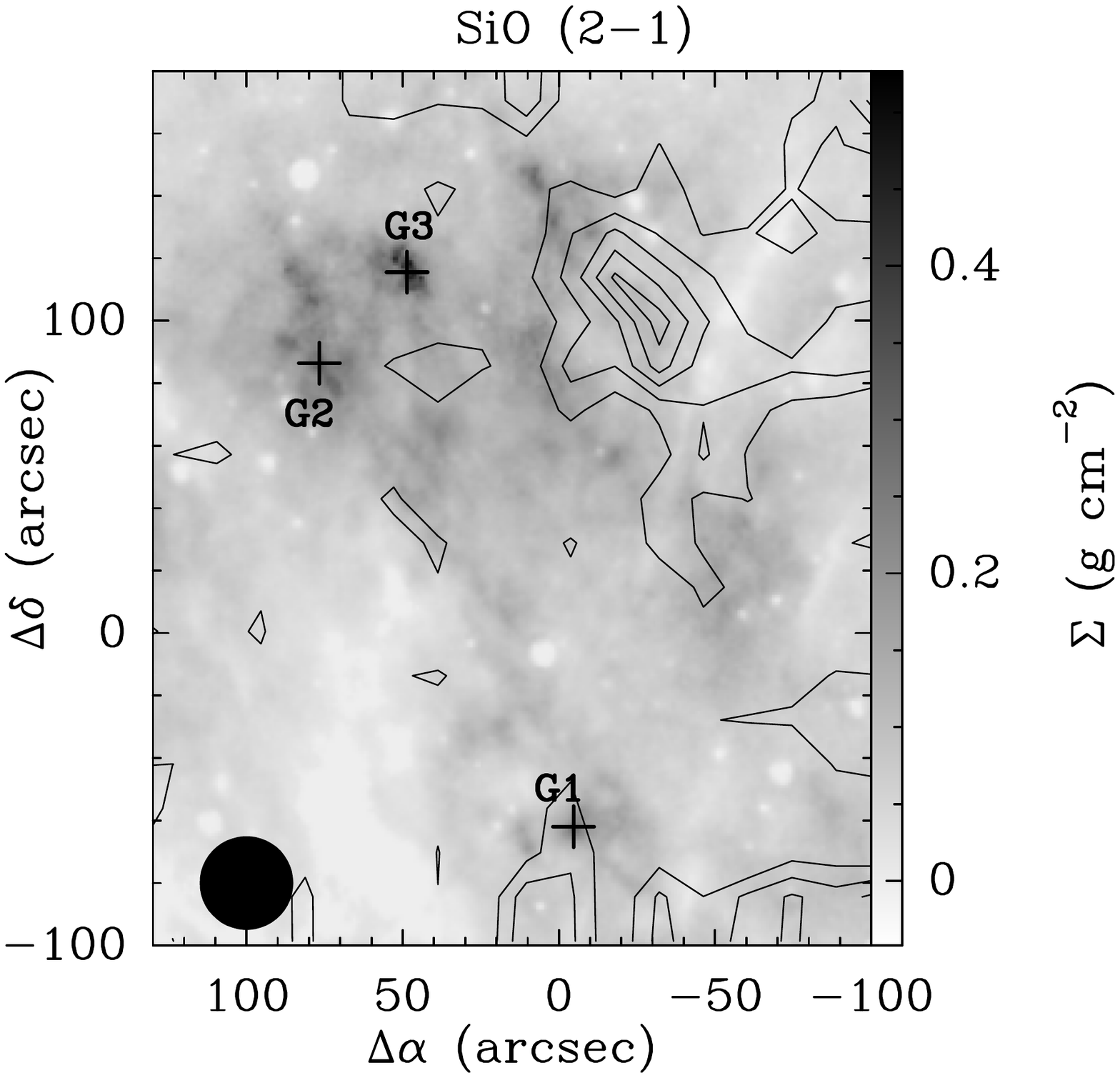}
\caption{Integrated intensity maps of the $\meth$(3-2) (left) and SiO (2-1) (right) lines toward cloud C (top panels), F (middle panels) and G (bottom panels). Emission levels (black contours) are 3$\sigma$, 6$\sigma$, 9$\sigma$, 12$\sigma$, 15$\sigma$, 21$\sigma$, 27$\sigma$, 33$\sigma$, 39$\sigma$, 45$\sigma$, 51$\sigma$ and 57$\sigma$ for $\meth$; 3$\sigma$, 6$\sigma$, 9$\sigma$, 12$\sigma$, 15$\sigma$, 18$\sigma$, 21$\sigma$, 24$\sigma$, 27$\sigma$ and 30$\sigma$ for SiO. The contours are superimposed on the H$_2$ mass surface density maps (gray scale) obtained by Kainulainen \& Tan (2013). \textit{Cloud C:} The integration ranges are 65 - 90 km s$^{-1}$ for SiO and 75 - 85 km s$^{-1}$ for $\meth$ and $\sigma$ = 0.20 K km s$^{-1}$ for both molecules. \textit{Cloud F:} The integration ranges are 40 - 80 km s$^{-1}$ for SiO and 52 - 64 km s$^{-1}$ for $\meth$. $\sigma$ = 0.24 and 0.25 K km s$^{-1}$ for SiO and $\meth$ respectively. \textit{Cloud G:} Integration ranges are 35 - 50 km s$^{-1}$ for SiO and 38 - 50 km s$^{-1}$ for $\meth$; $\sigma$ = 0.10 and 0.20 K km s$^{-1}$ for SiO and $\meth$, respectively. The core positions (black crosses; Butler $\&$ Tan 2009, 2012) and the beam sizes (black circles) are shown in all panels.}
\label{fig6}
\end{figure*}

\begin{figure*}
\centering
\includegraphics[scale=0.4,trim=0cm 2cm 0cm 1cm,clip=true]{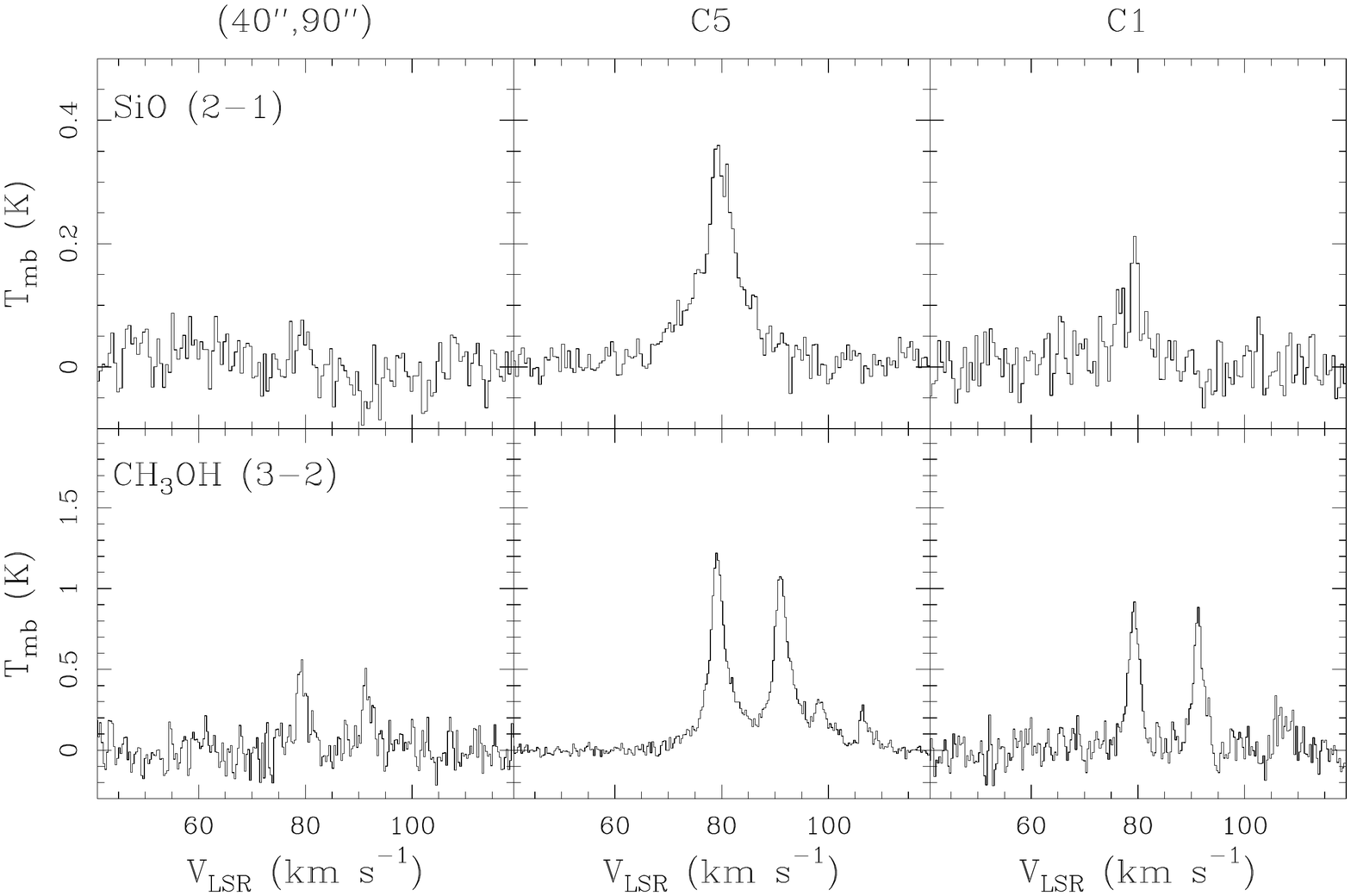}\\
\includegraphics[scale=0.4,trim=0cm 2cm 0cm 1cm,clip=true]{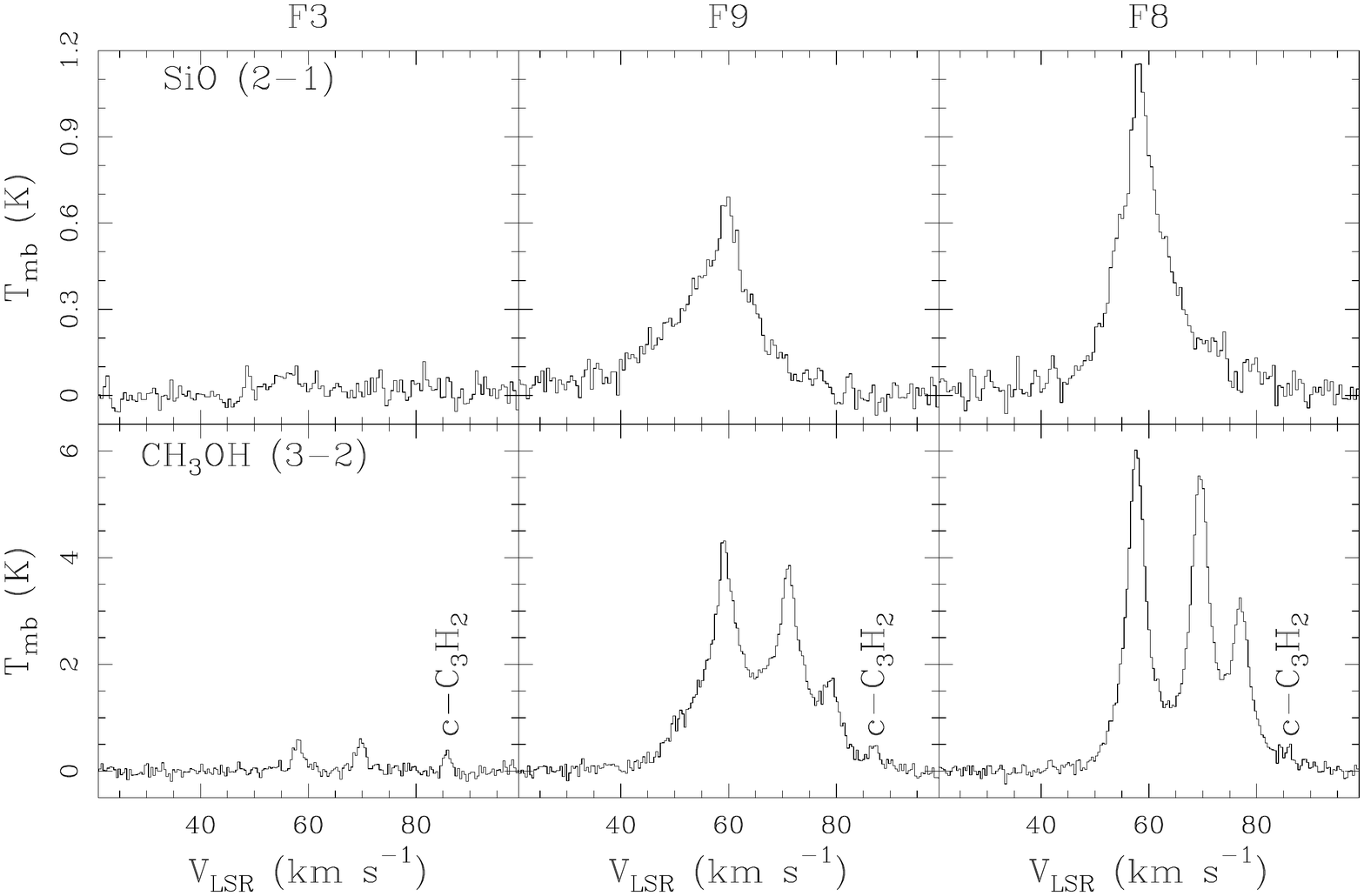}\\
\includegraphics[scale=0.4,trim=0cm 2cm 0cm 1cm,clip=true]{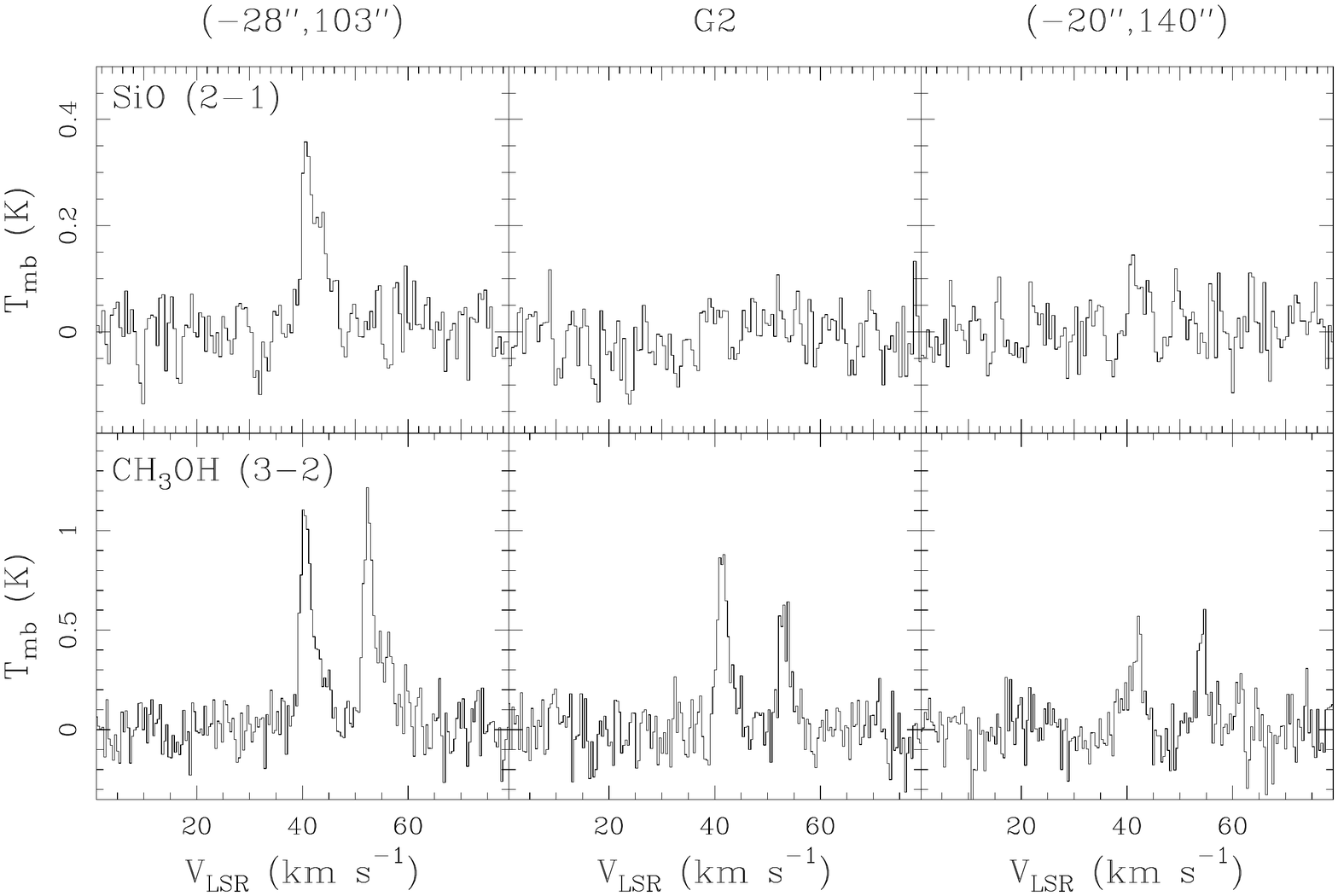}\\
\caption{SiO and $\meth$ spectra extracted toward three positions in cloud C (top panel), F (middle panel) and G (bottom panel). The corresponding position is indicated for every spectrum.}
\label{fig7}
\end{figure*}

In Figure~\ref{fig6}, the integrated intensity maps obtained for the shock tracers SiO and CH$_3$OH are shown. The emission levels for SiO (right panels) and $\meth$ (left panels) are superimposed on the H$_2$ mass surface density maps of \citet[][; in gray scale]{kainulainen2013}. Sample spectra extracted from different positions across clouds C (top panel), F (middle panel) and G (bottom panel) are also presented in Figure~\ref{fig7} to illustrate the change in line width and velocity across the clouds. 

For simplicity, we only show the integrated intensity map obtained for the 3$_{0,3}\rightarrow$2$_{0,2}$ $\meth$ transition.

\subsubsection{Shock tracer emission in clouds C and F}\label{shock_emission_C_F}

From Figure~\ref{fig6}, we find that the SiO and CH$_3$OH emission detected toward clouds C and F is bright and widespread over several parsecs (spatial extent of 4.8 pc$\times$6.8 pc for SiO and 6.3 pc$\times$6.8 pc for $\meth$ toward cloud C, and spatial extent of 2.3 pc$\times$4.8 pc for both SiO and $\meth$ toward cloud F). For cloud C (top panels), SiO is mainly associated with the massive cores in the cloud, especially those located along the central ridge known to be actively forming massive stars \citep[see][]{wang2011, zhang2014}. The $\meth$ emission is also bright toward the same regions where SiO is detected, although it shows an additional extension toward the northern quiescent parts of cloud C such as the emission toward C8 and the emission between C5 and the northern, very bright, infrared source (see 3$\sigma$ contour in the top left panel), following the morphology of the IRDC, and which remains unseen in SiO. 

For cloud F (middle panels), the emission of both SiO and $\meth$ shows a similar pattern to that observed in cloud C. The brightest emission peaks of SiO and $\meth$ coincide with the location of two active massive star-forming cores F8 and F9 \citep[see][]{rathborne2009, rathborne2011}, while the third SiO and CH$_3$OH peak, at offset (-35\arcsec,-20\arcsec), is just 0.43 pc away from the low-mass star-forming cluster identified by \cite{foster2014}, offset (-17\arcsec, -40\arcsec). It is then very likely that this third SiO and $\meth$ emission peak is associated with this star-forming cluster. As with cloud C, note that an additional extension is detected in $\meth$ with respect to SiO toward the central regions of cloud F (see 3$\sigma$ contour in the middle left panel), coinciding with the most quiescent regions and cores in this cloud, such as core F2.

\subsubsection{The case of cloud G}\label{shock_emission_G}
Very peculiar is the spatial distribution of the SiO and $\meth$ emission in cloud G (Figure~\ref{fig6}, bottom panels). Unexpectedly, the SiO emission detected toward this cloud appears concentrated toward the north-west at a location off the cloud with a visual extinction of just A$_v$$\sim$20-30 mag [at offset (-28\arcsec, 103\arcsec)], corresponding to a mass surface density of 0.06-0.09 g cm$^{-2}$, and far away from the massive cores reported in this cloud. This is also true for the brightest emission of $\meth$ detected in cloud G [offset (-15\arcsec, 110\arcsec)], as if both molecules had been enhanced by the same mechanism toward this position. We note however that, in contrast to SiO, $\meth$ is extended across the whole cloud and spatially follows the most quiescent regions in the IRDC (see Figure~\ref{fig6}). Therefore, the spatial distribution of the shock tracers SiO and $\meth$ in cloud G shows that {\it their emission peaks are completely offset from the known massive cores in the cloud, and that the SiO emission does not follow the morphology of the dense gas.} 

\subsubsection{The broad and narrow components}\label{shock_linewidth}
\begin{figure*}
\centering
\includegraphics[scale=0.8,trim= 0.2cm 6cm 0cm 5cm,clip=true]{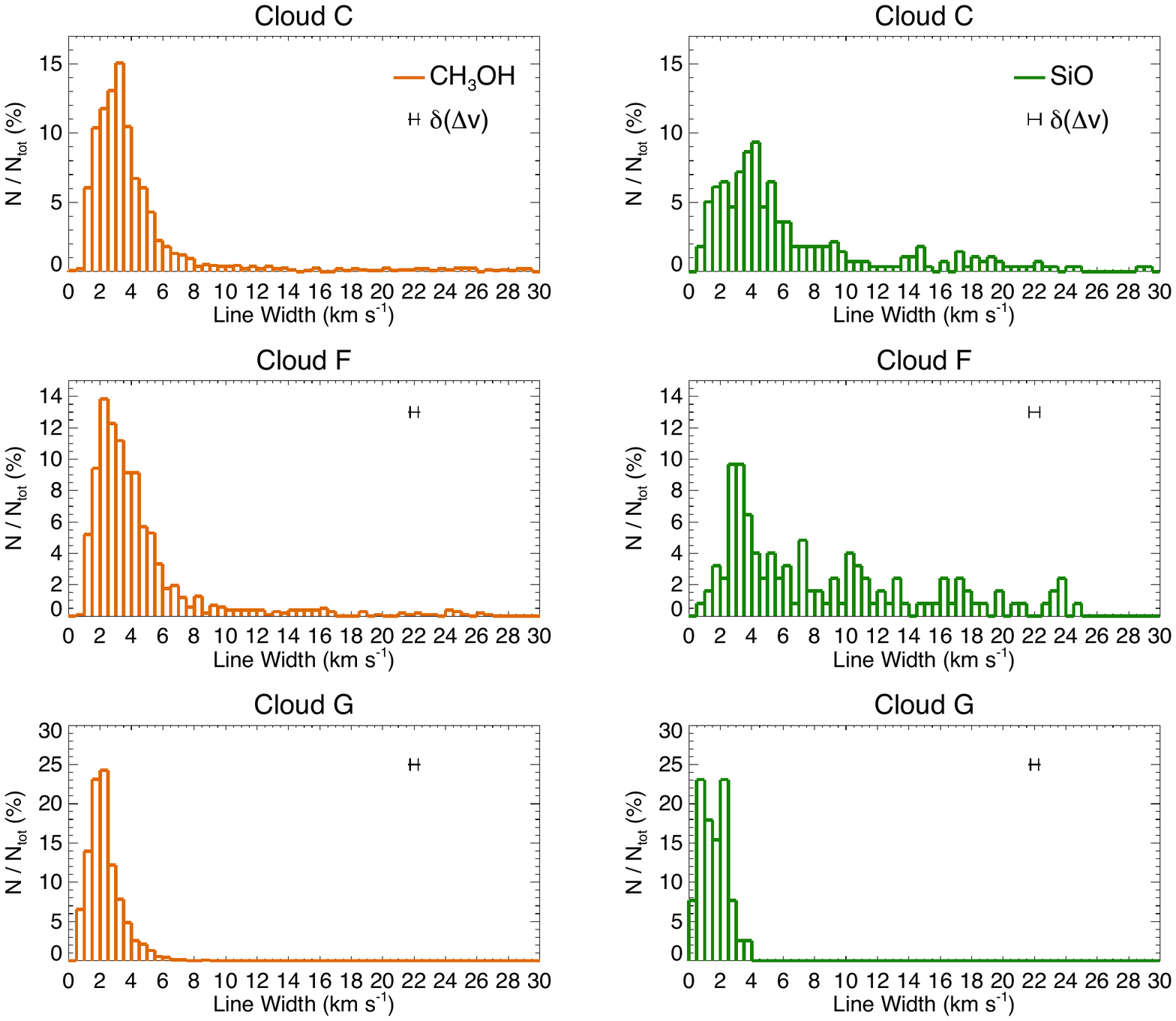}
\caption{Line width distributions of the $\meth$ (in orange) and SiO (in green) emissions obtained for clouds C (top panels), F (middle panels) and G (bottom panels). The histograms show the percentage of emission lines having line widths falling within each bin. Bin size for the $\meth$ and SiO distributions are 0.5 km s$^{-1}$ corresponding to 1/3 of the mean intensity-weighted linewidth obtained for the dense gas tracers. In each panel the mean uncertainty in the linewidths from the single fittings is indicated.}
\label{fig8}
\end{figure*}

By using SCOUSE and following a procedure similar to that performed for the dense gas tracers, in Figure~\ref{fig8} we present the line width distribution for SiO (in green) and $\meth$ (in orange) toward clouds C (top panels), F (middle panels) and G (bottom panels). The bin size in the x-axis is 1/3 of the mean intensity-weighted linewidth obtained for the dense gas tracers (i.e. 0.5 km s$^{-1}$). This allows to have a direct comparison between the distributions of SiO/$\meth$ and $\hmol$/$\hion$. In the y-axis we plot the percentage of emission lines having line widths falling within the bin size. For each histogram, we report the mean uncertainty on the linewidth, derived by averaging the uncertainty in all the linewidths obtained by SCOUSE. The $\meth$ histograms have been obtained considering the contributions from the transitions 3$_{0,3}\rightarrow$2$_{0,2}^{++}$, 3$_{-1,3}\rightarrow$2$_{-1,2}$ and 3$_{0,3}\rightarrow$2$_{0,2}$, which have been detected in all clouds. The fraction of velocities in each bin has been obtained considering the velocities obtained for the three transitions and then dividing by the total number of positions (i.e. the sum of the number of significant positions for each transition).

From Figure~\ref{fig8}, we find that the line width distributions of both SiO and $\meth$ for clouds C and F, show two main contributions: a narrow component with a well-defined peak at line widths between 2 and 4 km s$^{-1}$, and a more distributed broader component with line widths up to 30 km s$^{-1}$. In contrast to clouds C and F, the line width distribution obtained for cloud G only shows the narrow component for both SiO and $\meth$ with typical linewidths $\leq$ 3 km s$^{-1}$. From the line width distributions in Figure~\ref{fig8} only, it is difficult to define a clear threshold between the two components. Hence, we use the line width distribution of the high-density tracers $\hion$ and $\hmol$ to define such as threshold (see Figure~\ref{fig5}), since their emission is not affected by shocks. Since the linewidths from $\hion$ and $\hmol$ are narrower than 5 km s$^{-1}$, we thus establish the threshold between narrow and broad SiO and $\meth$ components at 5 km s$^{-1}$. In Table~\ref{tab5}, we report the percentage of SiO and $\meth$ lines having $\Delta v$ $<$ 5 km s$^{-1}$.

\begin{table}
\centering
\caption{Frequency of detection, in percentages, of the SiO and $\meth$ narrow emission towards clouds C, F and G, for thresholds $<$3 and 5 km s$^{-1}$}.
\begin{tabular}{cccc|ccc}
\hline
\hline
& \multicolumn{3}{c}{$<$ 3 km s$^{-1}$} & \multicolumn{3}{c}{$<$ 5 km s$^{-1}$}\\
\hline
Cloud   &C ($\%$) &F ($\%$)  &G ($\%$) & C ($\%$) &F ($\%$) &G ($\%$)\\
\hline
SiO     &28.9  	  &25.8      &97.4     &62.1      &45.2     &100\\
$\meth$ &47.7     &46.7      &83.4     &81.0      &85.6     &99.9\\
\hline
\end{tabular}
\label{tab5}
\end{table}

From Table~\ref{tab5}, it is clear that almost 100\% of the SiO and $\meth$ emission in cloud G is narrow, and therefore this component dominates the shock tracer emission in this cloud. On the other hand, the narrow emission in clouds C and F accounts for 45-60$\%$ of the SiO emission and 80-85$\%$ of $\meth$ measured in these clouds. We note that this difference is even greater when considering a threshold of 3 km s$^{-1}$, for which the fraction of SiO and $\meth$ narrow emission in cloud G is $\geq$83\% while it accounts for 25-50\% in clouds C and F. Hence, we conclude that while two line width components (narrow and broad) are clearly identified in clouds C and F, only the narrow component is observed in cloud G. This is clear in Figure~\ref{fig7} where SiO and $\meth$ spectra, extracted across the three clouds, are shown. In cloud C and F, the SiO and $\meth$ emission has linewidth $\geq$ 10 km s$^{-1}$ in positions corresponding to the active cores (see C5 in cloud C and F9 in cloud F) while in the more quiescent regions (see offset (40\arcsec,90\arcsec) in cloud C and core F3 in cloud F) the emission linewidth is as narrow as 5 km s$^{-1}$ or not detected. In cloud G (bottom right panels), the very broad component found in cloud C and F is not detected and even toward the SiO and $\meth$ peaks [see offset (-28\arcsec, 103\arcsec)] the emission is $\leq$ 4 km s$^{-1}$.

\subsubsection{Velocity Distribution}\label{shock_velocity}
\begin{figure*}
\centering
\includegraphics[scale=0.8,trim= 0cm 3cm 0cm 3cm,clip=true]{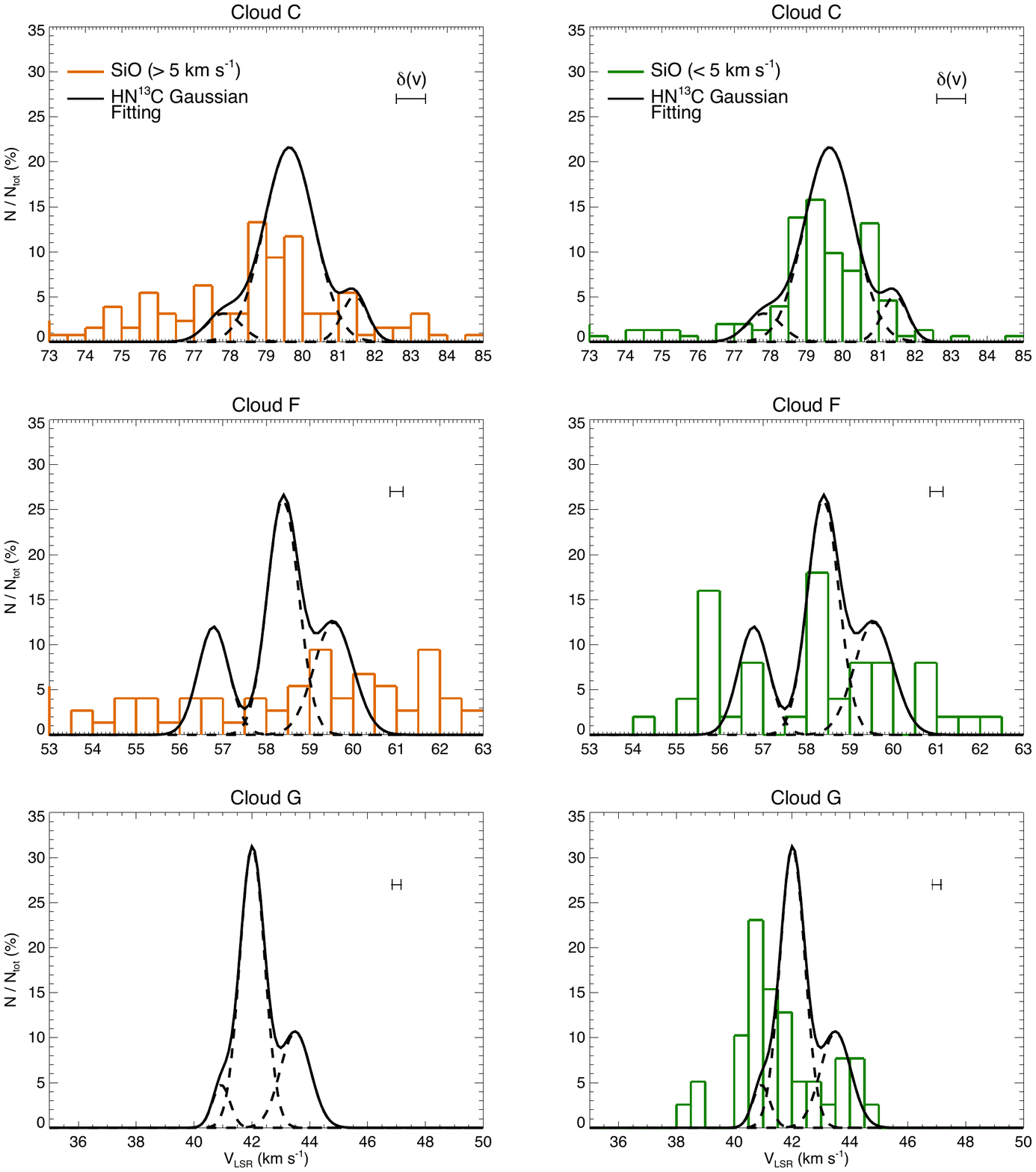}
\caption{Velocity distribution of the SiO broad (with linewidths $>$5 km s$^{-1}$, orange in left panels) and narrow (with linewidths $<$5 km s$^{-1}$, green in right panels) emission obtained for cloud C (top panels), F (middle panels) and G (bottom panels). The histograms show the percentage of emission lines having radial velocities falling within each bin, where the bin size is 0.5 km s$^{-1}$ for the three clouds, corresponding to 1/3 of the mean intensity-weighted linewidth obtained for the dense gas tracers. In each panel, the mean uncertainty for the central velocities of the single fittings is indicated. Black curves correspond to the gaussian fitting obtained for $\hmol$ in Section~\ref{dense_gas_filaments} (see also Figure~\ref{fig3}).}
\label{fig9}
\end{figure*}

\begin{figure*}
\centering
\includegraphics[scale=0.8,trim= 0cm 3cm 0cm 3cm,clip=true]{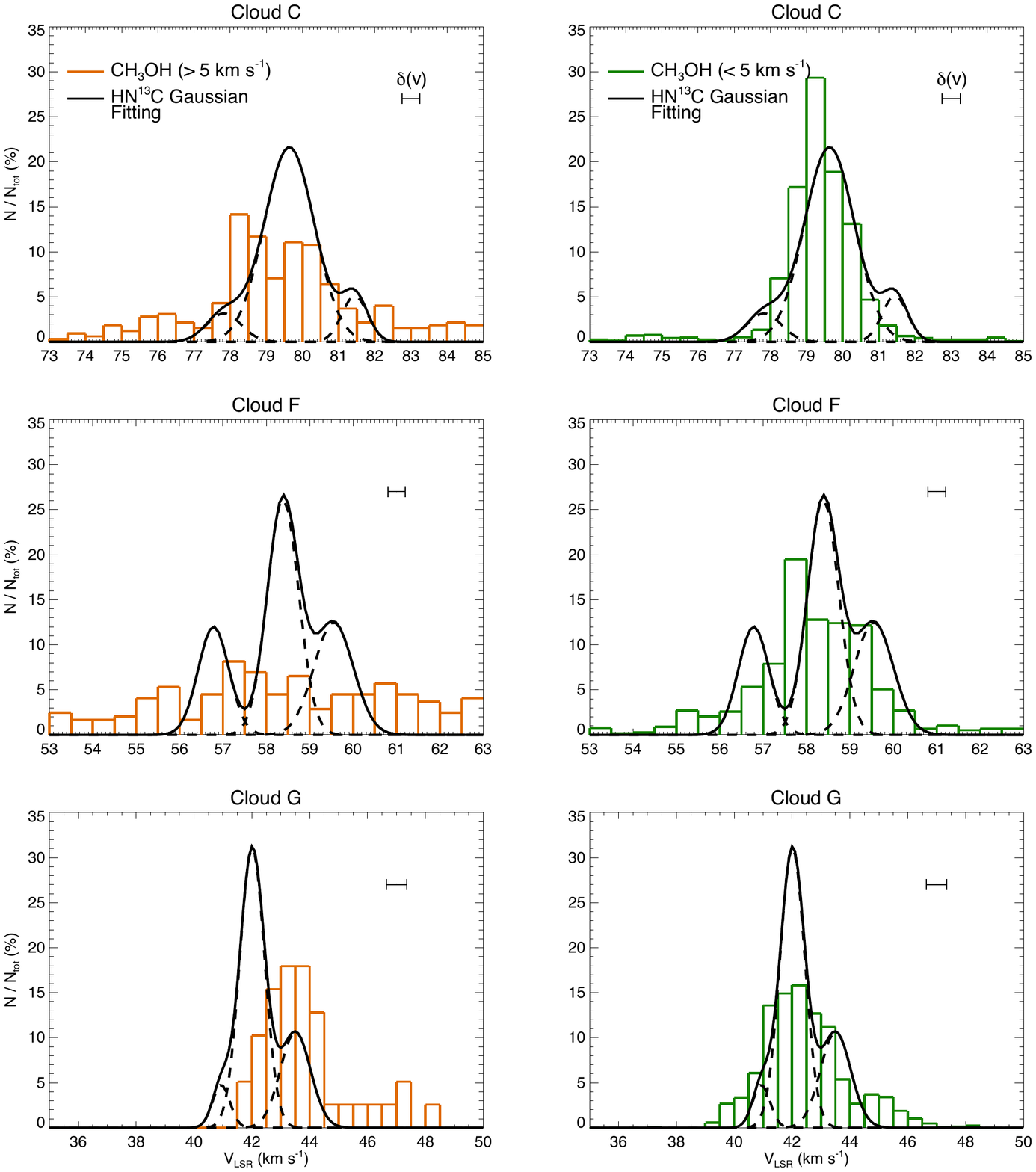}
\caption{Velocity distribution of the $\meth$ broad (linewidths $>$5 km s$^{-1}$, orange in left panels) and narrow (linewidths $<$5 km s$^{-1}$, green in right panels) emission obtained for clouds C (top panels), F (middle panels) and G (bottom panels). The histograms show the percentage of emission lines having radial velocities falling within each bin, with bin sizes 0.5 km s$^{-1}$ for the three clouds, corresponding to 1/3 of the mean intensity-weighted linewidth obtained for the dense gas tracers. The contribution from all $\meth$ transitions detected in each cloud has been considered. In each panel, the mean uncertainty for the central velocities of the single fittings is indicated. Black curves correspond to the gaussian fitting obtained for $\hmol$ in Section~\ref{dense_gas_filaments} (see also Figure~\ref{fig3}).}
\label{fig10}
\end{figure*}

We now investigate the distribution of the radial velocities of the two linewidth components (broad and narrow) detected in SiO and $\meth$ toward the three clouds. Figures~\ref{fig9} and \ref{fig10} present the velocity distribution of the broad (orange, left panels) and narrow (green, right panels) emission of, respectively, SiO and $\meth$. We compare these velocity distributions with the gaussian fitting obtained for the high-density tracer $\hmol$ (black curves; see also Figure~\ref{fig3}). The velocity distribution of the narrow (broad) component is divided by the number of positions in which the total narrow (broad) emission has been detected.

From Figures~\ref{fig9} and \ref{fig10}, it is clear that the SiO and $\meth$ broad components in clouds C and F show significant emission distributed across a wide range in central velocities. Furthermore, the velocity distributions are clearly unrelated to the velocity sub-structure identified from the dense gas (as probed by $\hion$ and $\hmol$), and they do not show any coherent sub-structures by themselves. For cloud G, however, the SiO emission does not show any broad component at all, while the $\meth$ broad component corresponds to just 0.1 $\%$ of the total emission and its line width is always $<$ 6 km s$^{-1}$. 

As shown in Figures~\ref{fig9} and \ref{fig10}, the narrow component is detected in all clouds and, in general, it arises mainly from the main (brightest) sub-structures detected in $\hion$ and $\hmol$. This indicates that the narrow component is moving as a whole at the central velocity of the corresponding cloud. One exception is found for the SiO narrow emission toward cloud G (right bottom panel in Figure~\ref{fig9}), for which the emission appears skewed to blue-shifted velocities. Note that the SiO velocity distribution for the narrow component in cloud F also shows a blue-shifted peak centered at $\sim$56 km s$^{-1}$ (see right middle panel in Figure~\ref{fig9}). This component is associated to core F8 located toward the south of cloud F (see Figure~\ref{fig6}), whose dense gas tracer emission also peaks at this blue-shifted velocity. However, unlike the narrow, blue-shifted SiO emission in cloud G, the SiO line profiles toward core F8 show both narrow and broad velocity components (see Figure~\ref{fig7}), indicating that the kinematics of the SiO gas in core F8 are driven by active star formation activity.

\section{Abundance of S\lowercase{i}O and CH$_3$OH in IRDCs}\label{abundances}
In this section, we estimate the abundance of the shock tracers SiO and $\meth$ measured for the narrow and broad components in the three clouds. The presence of several transitions allowed us to perform a multi-line excitation analysis for $\meth$ and estimate the excitation temperature, T$_{ex}$, and total column density, N$_{tot}$, for this molecule. In order to evaluate these quantities we re-gridded the $\meth$ data to the largest beam size used in our observations (30\arcsec) and extract spectra toward four positions in each cloud. The selected positions show either the broad or the narrow component and are representative of both the star-forming and quiescent gas. These positions are C1, C2, C8 and offset (40\arcsec, 90\arcsec) for cloud C; F2, F8, F9 and offset (-30\arcsec, -20\arcsec) for cloud F; and G1, G2, G3 and offset (-28\arcsec, 103\arcsec) for cloud G. Assuming the gas being in LTE conditions and the emission more extended than the beam size (filling factor$\sim$1), we used the new software {\sc MADCUBAIJ} \citep{rivilla2016} to calculate T$_{ex}$ and N$_{tot}$ of the narrow $\meth$ emission toward the selected positions. The derived $T_{ex}$ are in the range 6 - 11 K for all the clouds (see Table~\ref{tab6}), similar to those obtained from N$_2$H$^+$ and CO observations \citep[see e.g.][] {henshaw2014,jimenezserra2014}. The estimated N$_{tot}$($\meth$) lie in the range 3.5$\times$10$^{13}$ - 5.8$\times$10$^{14}$ cm$^{-2}$ for cloud C, 3.2$\times$10$^{13}$ - 2.4$\times$10$^{15}$ cm$^{-2}$ for cloud F and 9.8$\times$10$^{13}$-1.9$\times$10$^{14}$ cm$^{-2}$ for cloud G. 

A similar analysis is not possible for SiO, since only one transition was covered in our observations. Hence, in order to estimate the total column density of this molecular species we assumed the T$_{ex}$ derived from $\meth$ in each position and used MADCUBAIJ to calculate the corresponding N$_{tot}$(SiO) (see Table~\ref{tab6}). Toward cores F2, C8, G1, G2 and G3, no SiO emission was detected. Hence, only 3$\sigma$ upper limits to the abundance of this molecule have been estimated. The obtained N$_{tot}$(SiO) are in the range $\leq$3.0$\times$10$^{11}$ - 7.4$\times$10$^{12}$ cm$^{-2}$ for cloud C, $\leq$2.5$\times$10$^{11}$ - 5.4$\times$10$^{13}$ cm$^{-2}$ for cloud F and $\leq$2.9$\times$10$^{11}$-3.3$\times$10$^{12}$ cm$^{-2}$ for cloud G.

In order to evaluate the enhancement of SiO in the narrow component of clouds C, F and G, we use the emission of $\hion$ as a probe of the H$_2$ column density and calculate the column density ratio N(SiO)$/$N($\hion$). The abundance of narrow SiO and $\meth$ is estimated as follows:  

\begin{equation}
\chi(SiO) = \frac{N_{tot}(SiO)}{N_{tot}(\hion)} \times \frac{^{13}C}{^{12}C} \times \chi(HCO^+), 
\label{eq4}
\end{equation}

\begin{equation}
\chi(CH_3OH) = \frac{N_{tot}(CH_3OH)}{N_{tot}(\hion)} \times \frac{^{13}C}{^{12}C} \times \chi(HCO^+)
\label{eq5}
\end{equation}

\noindent
We assume a constant HCO$^+$ fractional abundance of $\sim$10$^{-8}$ \citep{irvine1987} and a $^{12}$C/$^{13}$C ratios of 40.2, 46.8 and 49.8 for clouds C, F and G respectively \citep{zeng2017}. The $\hion$ column densities are calculated following the same method as that used for SiO (see above). 
In the selected positions, the narrow emission from SiO and $\meth$ and the $\hion$ emission show similar central velocities and linewidths.
$\chi$(SiO) and $\chi$($\meth$) of the narrow component range between 6.3$\times$10$^{-11}$-7.4$\times$10$^{-10}$ and 0.9$\times$10$^{-8}$-5.8$\times$10$^{-8}$ for cloud C, $\leq$3.8$\times$10$^{-11}$-2.1$\times$10$^{-9}$ and 7.8$\times$10$^{-9}$-6.4$\times$10$^{-8}$ for cloud F, and $\leq$3.2$\times$10$^{-11}$-9.6$\times$10$^{-10}$ and 1.1$\times$10$^{-8}$-5.5$\times$10$^{-8}$ for cloud G.

The $\meth$ and SiO emission detected toward the active cores F8, F9, C1 and C2, also shows broad line profiles. Hence, we evaluated the column densities of the SiO and $\meth$ broad components observed toward these positions. We assumed T$_{ex}$ = 50 K, typical of shocked gas in molecular outflows (e.g. Jim\'enez-Serra et al. 2005), and we calculated the corresponding N$_{tot}$ using {\sc MADCUBAIJ}. Since no broad $\hion$ emission has been detected, we consider the 3$\sigma$ upper limits to the  N$_{tot}$($\hion$) to calculate the lower limits to the abundance of SiO and CH$_3$OH using Eqs~\ref{eq4} and ~\ref{eq5}. The total column densities of the broad SiO and $\meth$ components are listed in Table~\ref{tab6} and corresponds to $\chi$(SiO) and $\chi$($\meth$) in the range $\geq$9.5$\times$10$^{-10}$-2.0$\times$10$^{-9}$ and $\geq$1.6$\times$10$^{-8}$-9.7$\times$10$^{-8}$ for cloud C, and $\geq$8.7$\times$10$^{-9}$-9.3$\times$10$^{-9}$ and $\geq$1.2$\times$10$^{-7}$-5.6$\times$10$^{-7}$ for cloud F. A similar analysis is not possible in cloud G since no broad component has been detected toward this cloud.

From Table~\ref{tab6}, we find that the SiO/$\hion$ column density ratio for the narrow component is enhanced toward the active core positions in clouds C (e.g. C2) and F [e.g. F8, F9 and offset (-35\arcsec, -20\arcsec)] by a factor of $\geq$5 with respect to the upper limits measured in cores C8 and F2. In cloud G, the narrow SiO emission is enhanced by more than a factor of 10 toward its peak position, with respect to the upper limits measured in the core positions G1, G2 and G3, which implies that the $\chi$(SiO) toward the SiO peak in this cloud is enhanced by a factor of $\sim$10 with respect to the quiescent gas. 

\begin{table*}
\caption{Column densities, ratios and abundances of SiO, $\meth$ and $\hion$ measured for the narrow and broad components in several positions toward the three clouds. NC = Narrow Component; BC = Broad Component.}
\centering
\begin{tabular}{lllllllllllllllll}
\hline
\hline
&  & \multicolumn{2}{c}{N($\hion$)} & \multicolumn{2}{c}{N(SiO)} & \multicolumn{2}{c}{N($\meth$)} & \multicolumn{2}{c}{\multirow{2}{*}{$\frac{N(SiO)}{N(\hion)}$}} &  \multicolumn{2}{c}{\multirow{2}{*}{$\frac{N(\meth)}{N(\hion)}$}} & \multicolumn{2}{c}{$\chi_{SiO}$} & \multicolumn{2}{c}{$\chi_\meth$}\\

& & \multicolumn{2}{c}{($\times$10$^{12}$ cm$^{-2}$)} & \multicolumn{2}{c}{($\times$10$^{12}$ cm$^{-2}$)} & \multicolumn{2}{c}{($\times$10$^{14}$cm$^{-2}$)} & & & & &\multicolumn{2}{c}{($\times$10$^{-10}$)} & \multicolumn{2}{c}{($\times$10$^{-8}$)}\\
\hline
&Position & NC & BC & NC & BC & NC & BC & NC & BC & NC & BC & NC & BC & NC & BC\\
\hline
C &C1    &2.2 &$\leq$1.0-1.7 &0.6       &3.8       &1.3  &1.1      &2.5  &$\geq$3.8  &59.1  &$\geq$64.7   &0.6       &$\geq$9.5  &1.5 & $\geq$1.6\\
  &C2    &2.5 &$\leq$1.7     &7.4       &14        &5.8  &6.6      &3.0  &$\geq$8.2  &232.0 &$\geq$388.2  &7.4       &$\geq$20.0 &5.8 & $\geq$9.7\\
  &C8   &1.0 &$\cdots$      &$\leq$0.3 &$\cdots$  &0.3  &$\cdots$ &0.3  &$\cdots$   &34.0  &$\cdots$     &$\leq$0.8 &$\cdots$   &0.9 &$\cdots$\\
  &(40\arcsec, 90\arcsec)   &0.8 &$\cdots$      &0.4       &$\cdots$  &0.6  &$\cdots$ &0.5  &$\cdots$   &72.5  &$\cdots$     &1.2       &$\cdots$   &1.8 &$\cdots$\\
\hline
F &F2    &1.4 &$\cdots$      &$\leq$0.3 &$\cdots$  &0.5  &$\cdots$ &0.2  &$\cdots$   &37.1  &$\cdots$     &$\leq$0.4 &$\cdots$   &0.8 &$\cdots$\\
  &F8    &6.0 &$\leq$0.9-1.2 &8.0       &53.0      &18.0 &24.0     &1.3  &$\geq$44.2 &300.0 &$\geq$2696.6 &2.8       &$\geq$92.8 &6.3 &$\geq$56.6\\
  &F9   &3.3 &$\leq$1.3     &3.3       &54.0      &10.0 &7.7      &1.0  &$\geq$41.5 &303.0 &$\geq$592.3  &2.1       &$\geq$87.2 &6.4 &$\geq$12.4\\
  &(-35\arcsec, -20\arcsec) &0.6 &$\cdots$      &6.4       &$\cdots$  &1.6  &$\cdots$ &10.2 &$\cdots$   &254.0 &$\cdots$     &21.0      &$\cdots$   &5.3 &$\cdots$\\
\hline
G &G1    &1.8 &$\cdots$      &$\leq$0.3 &$\cdots$  &1.0  &$\cdots$ &0.2  &$\cdots$   &54.4  &$\cdots$	  &$\leq$0.3 &$\cdots$   &1.1 &$\cdots$\\
  &G2    &1.3 &$\cdots$      &$\leq$0.3 &$\cdots$  &1.0  &$\cdots$ &0.3  &$\cdots$   &76.9  &$\cdots$	  &$\leq$0.5 &$\cdots$   &1.5 &$\cdots$\\
  &G3    &0.9 &$\cdots$      &$\leq$0.3 &$\cdots$  &1.0  &$\cdots$ &0.4  &$\cdots$   &108.7 &$\cdots$	  &$\leq$0.7 &$\cdots$   &2.2 &$\cdots$\\
  &(-28\arcsec, 103\arcsec) &0.7 &$\cdots$      &3.3 &$\cdots$  &1.9  &$\cdots$ &4.8  &$\cdots$   &275.4 &$\cdots$     &9.6       &$\cdots$   &5.5 &$\cdots$\\
\hline
\end{tabular}
\label{tab6}
\end{table*}

\section{Gas Mass of the Sub-structures}
In this Section, we provide rough estimates of the H$_2$ gas masses of all the velocity-coherent sub-structures identified in Clouds C, F and G from the emission of the dense gas tracers (see Section 4.1). To do this, we first calculate the total column density of H$^{13}$CO$^{+}$ within each velocity-coherent component by using its derived mean peak intensity $\langle I_{peak}\rangle$, its mean intensity-weighted linewidth $\langle \Delta$v$\rangle$ and the number of emission lines fitted for each sub-structure, $n$.
 
We assume an excitation temperature \hbox{T$_{ex}$ = 9 K} for all components, as discussed in Section~\ref{abundances}. The inferred values of N($\hion$) are converted into H$_2$ mass surface densities by using (see also Hernandez et al. 2012):

\begin{equation}
\Sigma_{H_2} = 2 \times \mu_H \times \frac{^{12}C}{^{13}C} \times \frac{N(\hion)}{\chi(HCO^+)}
\end{equation}

\noindent
where the mass per H nucleus is $\mu_H$ = 2.34$\times$10$^{-24}$ gr, the assumed HCO$^+$ fractional abundance is $\chi$(HCO$^+$) = 10$^{-8}$ \citep{irvine1987} and the $^{12}$C/$^{13}$C isotopic ratios for clouds C, F and G are  40.2, 46.8 and 49.8 respectively \citep[see][]{zeng2017}. The H$_2$ gas mass of each velocity-coherent sub-structure is finally calculated by correcting by the pixel size area in units of cm$^{2}$. The obtained values are listed in Table~\ref{tab4}.

\begin{table*}
\centering
\caption{Mean peak intensity, mean intensity-weighted linewidth, number of fitted lines, $\hion$ total column density, H$_2$ gas mass surface densities and gas masses derived for all  velocity components identified in clouds C, F and G.}
\begin{tabular}{cccccccc}
\hline
\hline
& & $\langle I_{peak}\rangle$& $\langle \Delta$v$\rangle$ & $n$ & N($\hion$) & $\Sigma_{H_2}$ & M$_{H_2}$ \\
& & (K) & (km s$^{-1}$) & & ($\times$10$^{13}$ cm$^{-2}$) & (gr cm$^{-2}$) & (\sol)\\
\hline
Cloud C &Fil1  &0.24	&1.47  &58 	&2.0	&0.4	&227\\
	    &Fil2  &0.29	&2.05  &155	&8.8	&1.7	&988\\
		&Fil3  &0.21	&1.06  &30	&6.6	&0.1	&74\\
\hline
Cloud F &Fil1  &0.30	&1.31  &39	&1.5	&0.3	&109\\
		&Fil2  &0.50	&1.82  &94	&8.5	&1.9	&606\\
		&Fil3  &0.39	&1.44  &43	&2.4	&0.5	&172\\
\hline
Cloud G &Fil1  &0.25	&1.59  &30	&1.2	&0.3	&55\\
		&Fil2  &0.30	&1.87  &74	&4.1	&1.0	&192\\
		&Fil3  &0.22	&1.45  &44	&1.4	&0.3	&65\\
\hline
\end{tabular}
\label{tab4}
\end{table*}

From Table~\ref{tab4}, it is clear that the main velocity component in all clouds (Fil2) is always more massive (by factors $\sim$3-13) than their blue- (Fil1) and red-shifted (Fil3) components. This is in agreement with previous studies toward other filamentary IRDCs, which employed low-density gas tracers such as $^{13}$CO \citep{jimenezserra2014}. As expected, the dense gas in each velocity component represents only a small fraction of the total mass of the cloud (as compared to the total gas masses derived from near- and mid-IR extinction mapping by \citet{butler2012} and \citet{kainulainen2013}). A future study will establish the exact mass fractions of dense gas within these sub-structures by using large-scale maps of low-density gas tracers such as $^{13}$CO and C$^{18}$O obtained toward clouds C, F and G.

\section{Discussion}\label{discussion}
In the past decade, several scenarios have been proposed to explain the formation of filamentary IRDCs. These include the flow-driven formation scenario
and the cloud-cloud collision scenario. In the flow-driven scenario, rapid and warm atomic gas flows collide forming cold and filamentary structures due to thermal and gravitational instabilities \citep{vazquez2003, vanloo2007, hennebelle2008}. The collision velocities are of tens of km s$^{-1}$ and hence the emission of shock tracers (such as SiO and $\meth$) is expected to be very broad and the separation in velocities between sub-structures within the clouds are expected to be important \citep{hennebelle2008}. This first scenario presents major problems when a small magnetic field is considered \citep{kortgen2015,kortgen2016}. 

Alternatively, in the cloud-cloud collision scenario, already-molecular clouds collide at moderate velocities as a result of Galactic shear motions \citep[which are $\sim$10 km s$^{-1}$;][]{li2017}, and form filamentary structures that resemble IRDCs \citep{tan2000, tasker2009, vanloo2014, wu2015, wu2017a,wu2017b}. In this second scenario, the gas is already molecular and therefore fossil records of the cloud-cloud collision, such as high-J CO line emission or emission from molecular shock tracers such as SiO or CH$_3$OH, are expected to be detected \citep{jimenezserra2010, wu2015, hernandez2015, pon2015, pon2016a, pon2016b}. Since the shock velocities involved are low ($\leq$ 10 km s$^{-1}$) the fossil record emission is expected to be narrow, as opposed to the flow-driven scenario, where the high velocities involved in the flow collision would produce broader shock tracer emission.

From our results in Section~\ref{dense_gas_emission}, it is clear that the kinematic and physical structure of IRDCs is complex and shows multiple velocity-coherent structures separated in velocity space by just a few km s$^{-1}$ (ranging from 2.5 to 3.6 km s$^{-1}$; see Table~\ref{tab3}), and which are clearly associated with the IRDC structures detected in extinction. This resembles the physical structure and kinematics of the molecular gas observed toward the filamentary IRDC G035.39-00.33 \citep[already defined as cloud H;][]{henshaw2013, jimenezserra2014}. As with cloud H, the analysis of the emission of high-density gas tracers such as $\hion$ and $\hmol$ suggests that these velocity-coherent structures are connected in velocity space toward the positions where massive dense cores are found in these clouds (see Figure~\ref{fig4}), which would support the cloud-cloud collision scenario.     

In order to test this hypothesis, in Section~\ref{shock_emission} we have investigated the morphology and kinematics of the emission of two typical shock tracers, SiO and CH$_3$OH, to search for signatures of any possible cloud-cloud collision in clouds C, F and G. SiO and CH$_3$OH indeed are widespread in the three clouds with spatial extents of a few parsecs across. In addition, the kinematics of these shock tracers show that clouds C and F present two clear linewidth components, i.e. a broad component showing line widths up to 30 km s$^{-1}$ and a narrower component with line widths of $\leq$ 5 km s$^{-1}$ (see Figures~\ref{fig8}, ~\ref{fig9} and ~\ref{fig10}). In contrast, the SiO and CH$_3$OH emission from cloud G mainly shows narrow line profiles with a mean line width $<$ 2 km s$^{-1}$, i.e. even narrower than the emission found in clouds C and F. 
In the following, we discuss the possible origins of the two components observed in SiO and CH$_3$OH toward the three IRDCs analyzed in our study. 

\subsection{SiO emission in clouds C and F}\label{SiO_in_C_F}
Cloud F has been extensively studied in previous works \citep{rathborne2005,rathborne2008,foster2014} and evidence for ongoing star formation activity has been found toward several regions in the cloud. \cite{shepherd2004} and \cite{rathborne2005, rathborne2008} studied the massive active core F8 \citep[see][]{rathborne2006,butler2009} and classified it as a hot molecular core powered by an embedded B0 protostar. \cite{sanhueza2010} and \cite{sakai2013} found a hot core/outflow system associated with core F9. Furthermore, \cite{chambers2009} showed that most of the massive cores in the cloud have 24 $\mu m$, 8 $\mu m$ and 4.5 $\mu m$ emission; and \cite{foster2014}, by performing deep near-IR images in H- and K-bands with the Keck telescopes, reported the presence of a distributed population of low-mass protostars in cloud F (see for instance, the embedded young cluster driving a collection of outflows coincident with the SiO emission peak detected at offset [-35\arcsec,-20\arcsec]; see also Figure~\ref{fig6}). 

From all this, it is not surprising that the broad SiO component contributes to almost 50$\%$ of the total SiO emission measured in this cloud (see Table~\ref{tab5}). The SiO line profiles are clearly broad toward the most active cores in cloud F (see e.g. cores F8 and F9 in Figures~\ref{fig6} and ~\ref{fig7}), and toward the positions with weaker SiO emission the SiO line profiles clearly show a blue- or red-shifted broad component indicating that they are likely associated with outflow interaction (see offset (-35\arcsec, -20\arcsec) in Figure~\ref{fig7}).
This is supported by the fact that the abundances obtained for the SiO broad component are in agreement with those found in typical shocked outflowing gas \citep[of some 10$^{-9}$-10$^{-8}$; see][]{martinpintado1992, jimenezserra2005}.

Signatures of ongoing star formation activity are also present in cloud C. \cite{tan2016} and \cite{kong2017} found a bipolar outflow in $^{12}$CO (2-1) emission associated with core C1, while most of the other cores show emission at 24 $\mu m$, 8 $\mu m$ and 4.5 $\mu m$ \citep{chambers2009}. \cite{feng2016} also reported a molecular outflow seen in SiO 2$\rightarrow$1 emission toward core C1-S in this cloud, corresponding to a sub-structure in the C1 core of our sample. In fact, the SiO 2$\rightarrow$1 line profiles measured toward this outflow show a combination of a narrow component centered at the velocity of the quiescent gas, and a broad component that appears red-/blue-shifted by a few 10 km s$^{-1}$ with respect to the ambient gas. As proposed by \cite{feng2016}, while the broad component is associated with the most evolved post-shocked gas (with abundances $>$10$^{-9}$, similar to those reported in Section~\ref{abundances}), the narrow SiO component would correspond to an early stage in the propagation of MHD shock waves characterized by the magnetic precursor \citep[see also][]{jimenezserra2004, jimenezserra2009}. The abundance estimated by \cite{feng2016} for this narrow component, is $\sim$5$\times$10$^{-12}$, i.e. $\sim$10 times lower than reported in Table~\ref{tab6} toward this core (C1). However, note that \cite{feng2016} use interferometric data for the abundance estimation and it likely misses a significant fraction of the extended narrow SiO emission detected within the IRAM 30m beam. It is therefore likely that the narrow and broad SiO emission detected in cloud C also arises from shocked gas in molecular outflows.

\subsection{SiO emission in cloud G}\label{SiO_in_G}
As shown in Section~\ref{shock_emission}, the SiO emission in cloud G only shows a narrow component having a mean line width of 1.6 km s$^{-1}$, whose emission peaks at a position off the IRDC [offset (-28\arcsec, 103\arcsec)] and far away (56\arcsec-167\arcsec, i.e. 0.8-2.3 pc) from the location of the massive cores detected in this cloud (see Figure~\ref{fig6}). In order to investigate the origin of this narrow SiO emission, we will investigate in detail this region of the cloud.

\begin{figure}
\includegraphics[scale = 0.4,trim=3cm 2cm 0cm 0cm, clip=true]{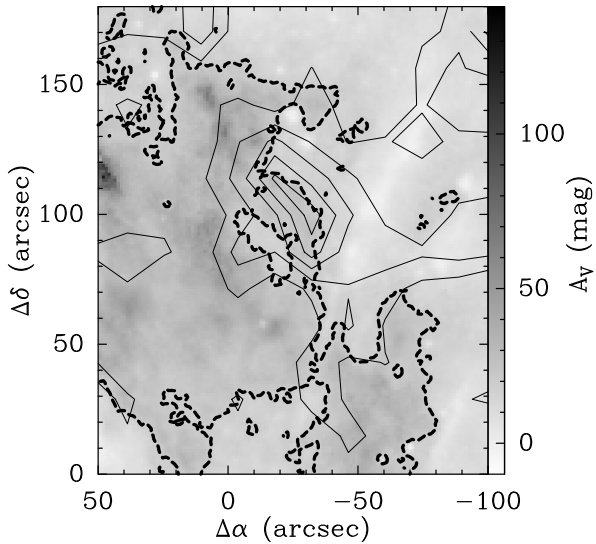}
\caption{Integrated intensity map of the SiO line emission observed toward cloud G (solid contours) and superimposed on the A$_v$ visual extinction map (gray scale) derived by Kainulainen \& Tan (2013). The velocity range for the SiO integrated intensity map is 35 - 50 km s$^{-1}$ and the contour levels are 3$\sigma$, 6$\sigma$, 9$\sigma$, 12$\sigma$ and 15$\sigma$ ($\sigma$ = 0.1 K km s$^{-1}$). Dashed contour corresponds to a visual extinction of A$_v$ = 20 mag.}
\label{fig11}
\end{figure}

In Figure~\ref{fig11}, we compare the morphology of the SiO emission in cloud G with that of the A$_v$ extinction map. With a dashed contour, we show the gas with dust extinctions higher than A$_v$ $\sim$ 20 mag. The comparison between the SiO emission and the dense gas in the cloud shows that SiO is extended and appears distributed mainly over the low extinction region in the map with A$_v$$\leq$ 20 mag. However, the peak emission of SiO interestingly follows the spatial distribution of a dense ridge with A$_v$$\geq$ 20 mag. 

\begin{figure*}
\centering
\includegraphics[trim= 1cm 7cm 1cm 2cm,clip=true,scale = 0.5]{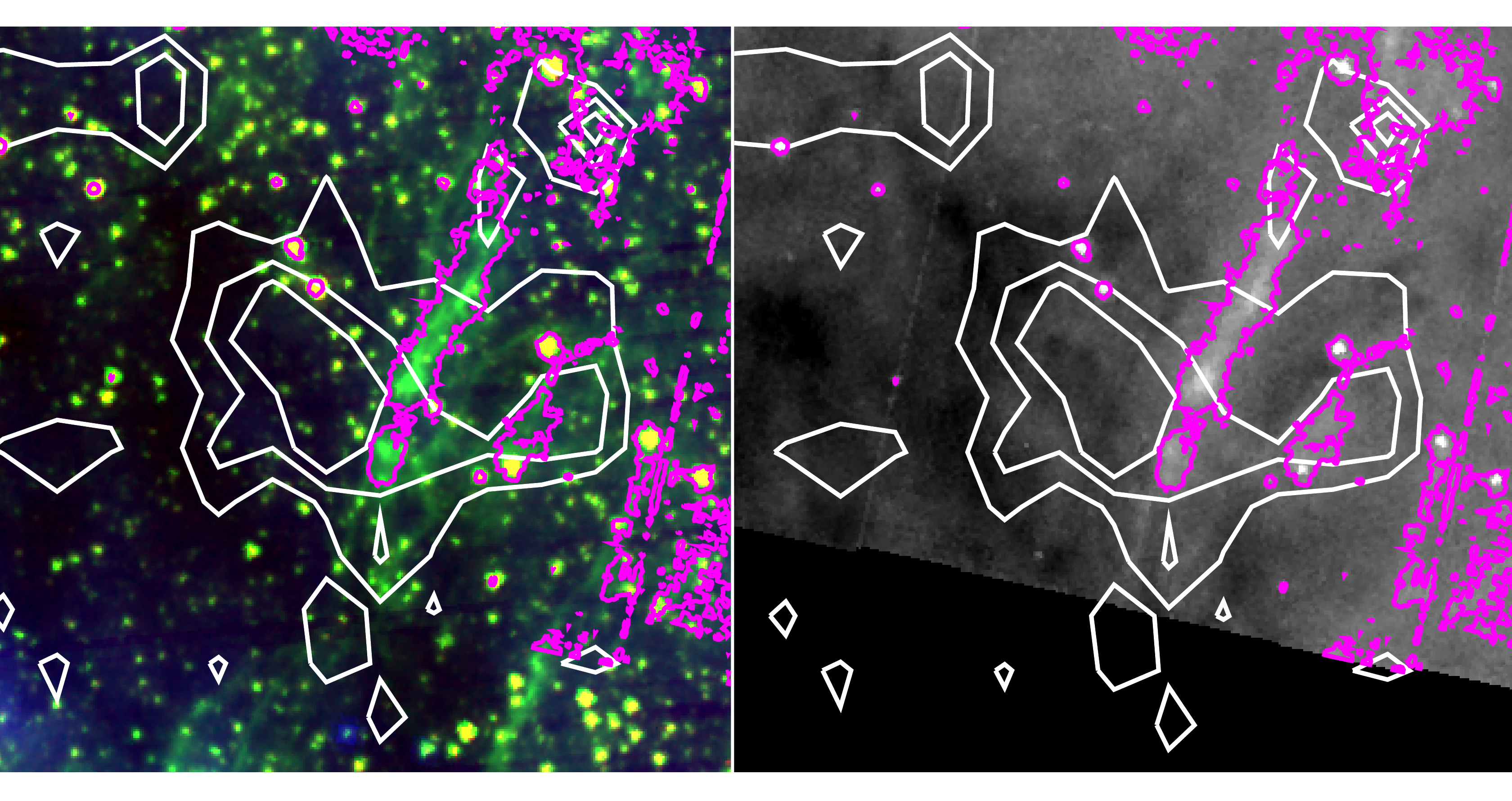}
\caption{Comparison of the SiO emission (white contours) with emissions at 3.6 $\mu$m, 4.5 $\mu$m, 8 $\mu$m and 24 $\mu$m, toward cloud G as obtained with the \textit{Spitzer Telescope}. \textit{Left panel: Spitzer} 3-color image with 3.6 $\mu$m emission displayed in red, 4.5 $\mu$m emission in green and 24 $\mu$m emission in blue. \textit{Right panel:} 8 $\mu$m emission toward cloud G displayed in gray scale. In both panels the SiO emission levels (white contours) correspond to 3$\sigma$, 6$\sigma$ and 9$\sigma$, with $\sigma$ = 0.1 K km s$^{-1}$. The 4.5 $\mu$m nebulosity is highlighted by the magenta contour corresponding to the 12$\sigma$ level at 4.5 $\mu$m and 100$\sigma$ level at 8 $\mu$m.}
\label{fig12}
\end{figure*}

In Figure~\ref{fig12}, we compare the SiO emission distribution (white contours) with the 8 $\mu$m emission distribution (right panel) and a 3-color image of the 3.6 $\mu$m (red), 4.5 $\mu$m (green) and 24 $\mu$m (blue) emission (left panel), all obtained with the \textit{Spitzer} Telescope. From Figure~\ref{fig12}, we find that most of the sources detected at 3.6 $\mu$m and 4.5 $\mu$m, and which overlap with the SiO emission, do not show any counterpart at 8 $\mu$m (only two 8 $\mu$m sources are detected toward the northern part of the SiO condensation). Emission at 4.5 $\mu$m is believed to be associated with shocked H$_2$ \citep{noriegacrespo2004} or CO gas in outflows \citep{marston2004} and it is considered as an indirect evidence of on-going star formation if coupled with 8 $\mu$m emission (typically associated with deeply embedded protostars). In addition, the 3.6 $\mu$m and 4.5 $\mu$m sources are mainly located outside the A$_v$ = 20 mag dense ridge and, therefore, it is unlikely that they represent young protostars driving outflows. Note that no 8 $\mu$m source is observed toward the location of the SiO/$\meth$ peak in cloud G.  

Figure~\ref{fig12} also shows very extended 4.5 $\mu$m emission running through the SiO extended emission from the north-west to the south-east (see magenta contours in both right and left panels). This emission is not associated with any 24 $\mu$m object (in blue in the left panel) or 8 $\mu$m source in the region (gray scale in right panel), and it does not seem to have originated from any embedded protostar. Instead, it appears as an extended nebulosity containing a collection of 4.5 $\mu$m point sources. For this reason, and because the SiO emission does not peak toward this nebulosity (Figure~\ref{fig12}), it is unlikely that this extended 4.5 $\mu$m emission represents an outflow. 

We speculate that the nebulous 4.5 $\mu$m emission is a remnant of a large-scale interaction. In this scenario, this interaction would have compressed the region towards the western edge of cloud G, yielding the A$_v$ = 20 mag ridge seen in extinction. The extended SiO emission would then be a fossil record of this interaction. This idea would be supported by the fact that the radial velocities of SiO toward this condensation do not peak at the same velocity as the high-density gas probed by HN$^{13}$C, but appear at blue-shifted velocities (see Figure~\ref{fig9}). Note that the histogram with the radial velocities of the narrow CH$_3$OH emission in cloud G also contains the contribution associated with the bulk of the IRDC and, as a result, it does not show the same pattern as the narrow SiO. The spread in velocities observed for SiO would support the cloud-cloud collision scenario, since it is expected from simulations of cloud-cloud collisions \citep{wu2017a,wu2017b}. in addition, the differences in the kinematics of the SiO emission and the high-density gas may indicate that the gas compressed in the large-scale interaction did not have enough time to relax to the velocities of the bulk of the gas in the IRDC. This interaction had to be relatively gentle since the SiO emission shows linewidths $\leq$ 3 km s$^{-1}$ (see Figure~\ref{fig8}), which are much narrower than the ones found in clouds C and F. 

Alternatively, the compression of the molecular gas in the A$_v$ = 20 mag dense ridge may have triggered the formation of a population of low-mass protostars. Bipolar outflows driven by the newly formed low-mass stars could be responsible for the presence of the detected SiO emission \citep{martinpintado1992,zhang2000}. In this scenario, the axis of the putative outflows would have to be close to the plane of the sky, with their red-shifted lobes being screened by the A$_v$ = 20 mag dense ridge. This configuration would explain the narrow SiO line profiles in cloud G that peak mostly at blue-shifted velocities. As recently shown by   
\citet{stephens2017}, such a configuration is highly unlikely since molecular outflows are found to be randomly oriented with respect to the molecular filaments in which they are embedded. In addition, projection effects are not expected to play a major role in the linewidth distribution of SiO in IRDCs, since the probability to detect outflows with broad emission assuming a random distribution of outflow inclination angles, is significantly lower than that to detect outflows with narrow SiO lines \citep[see Section 3.2 in][]{beuther2007}. This is in contradiction with observations. 

Narrow SiO emission has also been detected in very young outflows and proposed to be driven by the early stages of the propagation of magneto-hydrodynamic (MHD) shock waves \citep[i.e. via the magnetic precursor; see][]{jimenezserra2004,jimenezserra2005}. This kind of emission is however rare and it has only been reported toward two sources \citep{jimenezserra2004,feng2016}. Another possibility would be that narrow SiO originates from shocked gas that has been decelerated by the interaction with the surrounding molecular gas envelope (see Lefloch et al. 1998). In this scenario, the extension of narrow SiO emission detected in cloud G would require $\sim$70 outflow sources for which their associated shocked gas would have been decelerated by the interaction with the Av=20 mag-extinction ridge. Higher angular resolution observations are therefore needed to establish the origin of narrow SiO in cloud G.

Finally, we note that it is very unlikely that the SiO emission detected in cloud G is arising from the merging of the internal sub-structures found in the dense gas (see Section~\ref{dense_gas_filaments}, due to the fact that the separation in velocities is too low to account for the observed abundance toward the SiO peak. This is also supported by the fact that the sub-structures detected in cloud G do not spatially overlap in the region where the SiO emission is present.

\subsection{SiO in actively star-forming ridges: Comparison with previous works}\label{SiO_comparison}
The emission from the 2$\rightarrow$1 rotational transition of SiO has also been studied toward at least two very active high-mass star forming regions: W43 \citep{motte2003, nguyen2011, nguyen2013, louvet2014, louvet2016} and Cygnus X \citep{wendker1991, schneider2006, motte2007, duartecabral2014, duartecabral2016}. These regions are both classified as mini-sturburts due to their high level of star formation activity, and their extended emission of SiO has been proposed to have originated in large-scale interactions \citep{nguyen2013, louvet2016}. The SiO lines detected toward W43 show a single gaussian line profile with typical linewidths $\geq$ 6 km s$^{-1}$ \citep{nguyen2013, louvet2016} and a mean SiO total column density of 6$\times$10$^{13}$ cm$^{-2}$ has been estimated for this source \citep[see][]{nguyen2013, louvet2016}. A similar analysis has been performed on Cygnus X by \cite{motte2007} and \cite{duartecabral2014}. The authors identified two linewidth components in the SiO line profiles: a very broad emission with linewidths $\sim$17 km s$^{-1}$ and a narrower emission with linewidths in the range of 3-6 km s$^{-1}$. \cite{duartecabral2014} report a SiO total column density of 0.1-4$\times$10$^{12}$ cm$^{-2}$ toward Cygnus X. 

If we compare the properties of the SiO emission toward W43 and Cygnus-X with our results from cloud G, we find that the SiO emission detected in W43 shows broader line profiles than those from cloud G ($\geq$ 6 km s$^{-1}$ in W43 vs. $\leq$ 3 km s$^{-1}$ in cloud G; see Figure~\ref{fig8}), and the total column density in W43 is higher by a factor $\geq$20 with respect to that found in cloud G. On the other hand, the SiO linewidth and the total column density estimated toward Cygnus X are comparable with the results found in cloud G and presented in Table~\ref{tab6}. However, both W43 and Cygnus X show a level of star formation activity that is much higher than that of cloud G. This suggests that W43 and Cygnus X are in a more evolved evolutionary stage than cloud G and therefore, the SiO extended emission detected toward these sources may be affected/influenced by star formation. Cloud G may thus represent a unique case where the effects of large-scale interactions can be tested {\it before the onset of star formation.}

The SiO emission detected in W43 and Cygnus-X is comparable to our results from clouds C and F. Both the broad and narrow SiO components detected in the two clouds show linewidths similar to that in W43 and Cygnus-X and the SiO total column density in W43 is comparable to those listed in Table~\ref{tab6} for the two clouds of our sample. This similarity and the fact that all the four sources are hosting active sites of star formation would be an evidence of the common star-formation related origin of the SiO emission toward such sources.

\subsection{$\meth$ emission in IRDCs}\label{methanol_in_irdcs}
As with the SiO emission, the $\meth$ line profile shows a very broad linewidth component toward clouds C and F, the most active clouds of the sample. This broad component shows properties very similar to those of the broad SiO emission in the two clouds i.e. they show a similar spatial distribution and mean linewidth. It is hence likely that, as for SiO, the $\meth$ broad emission arises from post-shocked gas in molecular outflows \citep[see][]{jimenezserra2005, tafalla2010}. This is supported by the fact that the $\meth$ broad component represents less than 0.1\% of the total emission in cloud G, where the star formation activity is much lower.

As discussed in Section~\ref{shock_emission}, the $\meth$ emission also shows a very narrow linewidth component in all the three clouds. This narrow component is very extended in all of the three clouds and, in particular, in cloud G where it is found to only be narrow (see Section~\ref{shock_emission}). From the bottom panels of Figure~\ref{fig6}, both SiO and $\meth$ emission are enhanced toward the north-west region of cloud G and hence it is likely that, toward this region the two molecules are affected by the same physical process and thus have the same origin (either large-scale interaction or a young population of low-mass protostars; see Section~\ref{SiO_in_G}). The abundance of narrow CH$_3$OH derived toward this position is indeed a factor of $\geq$2 higher than toward the quiescent cores G1, G2 and G3. However, unlike SiO, $\meth$ shows narrow emission associated with the bulk of the IRDC, where no SiO is detected. Narrow $\meth$ toward the cores in cloud G cannot be due to star formation because these cores are quiescent and do not show any signature of on-going star formation \citep[see][]{chambers2009}. Hence, some additional mechanism has to be responsible for this emission. This is also true for clouds C and F, where narrow $\meth$ appears more extended than the narrow SiO component. The narrow emission of $\meth$ also peaks at the radial velocity of the clouds (see Figure~\ref{fig10}), indicating that most of this emission is related to the quiescent gas. 

Several mechanisms can account for the release of $\meth$ into the gas phase. These include UV-induced photo-desorption and chemical reactive desorption\footnote{In the process of chemical reactive desorption, the newly formed molecule possesses an energy surplus that allows it to evaporate \citep[see][]{minissale2016}.}. As reported by \cite{martindomenech2016}, UV-induced photo-desorption is very unlikely since UV photons dissociate $\meth$ into radicals forming smaller molecules such as CO, CO$_2$ and H$_2$CO that are then released into the gas phase. On the other hand, chemical reactive desorption allows the desorption of $\meth$ itself after its formation on grain surfaces via hydrogenation, representing a viable mechanism for the production of narrow $\meth$ in IRDCs not associated with star formation. This process also explains the presence of $\meth$ and other complex organics in the gas phase of pre-stellar cores such as L1498, L1517B and L1544 \citep{bizzocchi2014, jimenezserra2016,vasyunin2017}. The abundance of $\meth$ in the L1544 pre-stellar core is $\sim$6$\times$10$^{-9}$ \citep{vastel2014}, which is comparable to the $\meth$ abundances estimated toward clouds C, F and G in the quiescent gas for the narrow component (see Table~\ref{tab6}). 

The presence of internal shocks of a few km s$^{-1}$ can also be responsible for the narrow $\meth$ emission in IRDCs \citep{pon2015}. These shocks are due to virialized motions of the gas within the clouds and could possibly account for the estimated $\meth$ abundance. However, at the current spatial resolution it is not possible to resolve such internal shocks. We exclude the scenario of thermal desorption in the quiescent regions of clouds C, F and G because the typical temperatures of IRDCs \citep[$\leq$20 K;][]{pillai2006} are much lower than that required for $\meth$ to evaporate \citep{collings2004}. However, note that this desorption mechanism is possibly responsible for the narrow $\meth$ lines detected toward star-forming cores in clouds C and F because they likely arise from the hot envelopes around young protostars. 

\section{Conclusions}\label{conclusion}
Using the IRAM 30m single dish telescope, we performed a detailed analysis of the $\hion$, $\hmol$, SiO and $\meth$ emission toward three filamentary IRDCs: \irdcC, \irdcF\space and \irdcG\space (clouds C, F and G, respectively). We investigated the spatial distribution and kinematics of dense gas ($\hion$ and $\hmol$) and shock (SiO and $\meth$) tracers. We studied the line profiles for all the transitions, including studying the line width and central velocity distributions. The results we obtained can be summarized as follows:

\begin{itemize}

\item[i)] In all three clouds of our sample, the dense gas emission shows multiple velocity components corresponding to three independent sub-structures that are spatially and kinematically well defined. The brightest sub-structures probe the ambient gas moving at the central velocity of the clouds, while the red and blue shifted sub-structures have velocity drifts of $\sim$1 km s$^{-1}$ with respect to the central velocity of the clouds.

\item[ii)] The $\hion$ and $\hmol$ emission lines show similar line width, always $\leq$ 5 km s$^{-1}$. The mean line width values (in the range 1.8 -1.6 km s$^{-1}$) obtained for the two molecules are comparable in the three clouds. Furthermore, from the linewidth distributions shown in Figure~\ref{fig5} only one line width component can be identified. 

\item[iii)] The spatial distribution of shock tracers is widespread at a parsec scale in all the clouds. In clouds C and F the emission is also affected by the presence of several active cores, while in cloud G, the SiO and $\meth$ emission distributions are very different. In this cloud the SiO emission is concentrated around its peak, in the north east region, while the $\meth$ emission is distributed along the cloud and follows well its filamentary structure. In both cases, the emission distribution is not correlated with the (inactive) cores of the cloud.

\item[iv)] The shock tracers' emission shows multiple line width components: a narrow ($\leq$ 5 km s$^{-1}$), widespread component across the clouds that is present in almost every position with significant emission, and a broad (up to 30 km s$^{-1}$) component that is strictly related to the positions of the cores in the cloud. In clouds C and F, both components are detected. In cloud G, the SiO emission is only narrow, while for the $\meth$ emission, 0.1$\%$ of the total emission present lines broader than 5 km s$^{-1}$. However, this small fraction presents line width always $<$ 6 km s$^{-1}$ and is not significant compared to the total emission.

\item[v)] Neither the broad nor the narrow component in the SiO and $\meth$ emission follow the high-density sub-filamentary structure. Furthermore, the broad component is distributed along a wide velocity range, while the narrow component is less extended in velocity. Narrow SiO emission in cloud G shows a radial velocity distribution skewed to blue-shifted velocities.

\item[vi)] In clouds C and F, the shock tracers' emission is likely due to the ongoing star formation activity. The presence of outflows in these clouds would be responsible for both the broad and the narrow components, corresponding to the post-shocked gas and the early stage in the propagation of MHD shock waves, respectively.

\item[vii)] In cloud G, the very low level of star formation activity and the very narrow linewidth of the shock tracers emission suggest a different origin for the SiO and the enhanced $\meth$ emission toward the north-west of this cloud. Particularly, the SiO emission could be a fossil record of a previous large-scale interaction occurring nearby. As a second possibility, the mentioned large-scale interaction could have triggered the formation of a low-mass star population within a A$_v$=20 mag-extinction ridge, whose SiO emission originates from outflows. Such outflows should show an almost parallel orientation in the plane of the sky with their red-shifted lobes being screened by the A$_v$=20 mag ridge. 

\item[viii)] The narrow $\meth$ emission in cloud G is more extended than the corresponding SiO emission and shows velocities very close to the central velocity of the cloud. Hence, it is mainly associated with the quiescent gas and it is likely produced by the chemical-reactive desorption mechanism. 
\end{itemize}

\section*{Acknowledgements}\label{acknow}
We thank the anonymous referee for their comments which have helped to improve the clarity of this paper. I.J.-S. acknowledges the financial support received from the People Programme (Marie Curie Actions) of the European Union's Seventh Framework Programme (FP7/2007-2013) under REA grant agreement PIIF-GA-2011-301538, and from the STFC through an Ernest Rutherford Fellowship (proposal number ST/L004801). The research leading to these results has also received funding from the European Commission (FP/2007-2013) under grant agreement No 283393 (RadioNet3).
P. C. acknowledges support from the European Research Council (ERC; project PALs 320620).
Partial salary support for A. P. was provided by a Canadian
Institute for Theoretical Astrophysics (CITA) National
Fellowship.

\bibliographystyle{mn2e}\label{lastpage}
\bibliography{IRDCs_gc}

\end{document}